\documentclass[fleqn,usenatbib]{rasti}

\usepackage{newtxtext,newtxmath}
\usepackage{anyfontsize}

\usepackage[T1]{fontenc}

\DeclareRobustCommand{\VAN}[3]{#2}
\let\VANthebibliography\thebibliography
\def\thebibliography{\DeclareRobustCommand{\VAN}[3]{##3}\VANthebibliography}

\usepackage{graphicx}	
\usepackage{amsmath}	
\usepackage{subcaption}
\usepackage{float}

\title[GNSS PWV against 210~GHz WVR PWV at H.E.S.S.]{Analysis of the accuracy of GNSS inferred precipitable water vapour against that from a 210~GHz WVR at the H.E.S.S. site}

\author[L. Frans, M. Backes, H. Falcke and T. Venturi]{
Lott Frans$^{1}$\thanks{E-mail: lfrans@unam.na},
Michael Backes$^{1,4}$,
Heino Falcke$^{2}$,
and Tiziana Venturi$^{3}$
\\
$^{1}$Department of Physics, Chemistry $\&$ Material Science, University of Namibia, Private Bag 13301, Windhoek, Namibia\\
$^{2}$Department of Astrophysics, Institute for Mathematics, Astrophysics and Particle Physics, Radboud University, P.O. Box 9010, 6500 GL Nijmegen, The Netherlands\\
$^{3}$Istituto di Radioastronomia, Instituto Nazionale di Astrofisica, Via Gobetti 101, 40129 Bologna, Italy\\
$^{4}$Centre for Space Research, North-West University, Private Bag X6001, Potchefstroom 2520, South Africa
}

\date{Accepted 18/04/2025. Received 16/04/2025; in original form 29/12/2024}

\pubyear{\the\year{}}

\begin{document}
\label{firstpage}
\pagerange{\pageref{firstpage}--\pageref{lastpage}}
\maketitle

\begin{abstract}
The High Energy Stereoscopic System~(H.E.S.S.) site and the Gamsberg Mountain have been identified as potential sites for the Africa Millimetre Telescope~(AMT). The AMT is poised to observe at millimetre and possibly at submillimetre wavelengths. At these wavelengths, precipitable water vapour~(PWV) in the atmosphere is the main source of opacity during observations and therefore needs to be accurately assessed at the potential sites for the AMT. In order to investigate the PWV conditions for the AMT, identical Global Navigation Satellite System~(GNSS) stations were installed and used to assess the PWV at the two potential sites. In this study, the accuracy of those PWV measurements by the GNSS stations was assessed by comparing the H.E.S.S. installed GNSS station PWV measurements to that from a 210~GHz Water Vapour Radiometer~(WVR) also installed at the H.E.S.S. site. A correlation of 98\% and an offset of 0.34~mm was found between the GNSS station and the 210~GHz WVR PWV data when on-site pressure and the Nevada Geodetic Laboratory~(NGL) weighted mean temperature~($\mathrm{T_m}$) were used calculate the GNSS station PWV data. In comparison, the offset reduces to 0.15~mm when on-site derived $\mathrm{T_m}$ and pressure were used to calculate the GNSS station PWV. The results show that the GNSS station with on-site meteorological data can be used with high accuracy to reliably determine the PWV conditions at the H.E.S.S. site.
\end{abstract}




\section{Introduction}
The High Energy Stereoscopic System~(H.E.S.S.) site~\citep{hess_nam} is one of two potential sites for the Africa Millimetre Telescope~(AMT) with the Gamsberg Mountain being the other option~\citep{GNSS_lott, AMT_backes}. Both the H.E.S.S. site and the Gamsberg Mountain are located within the Khomas Highlands area of Namibia and are approximately 30~km apart. The area is known to have pristine conditions for astronomy and already consists of the H.E.S.S. observatory. The H.E.S.S. observatory observes gamma-rays and consists of an array of five imaging atmospheric Cherenkov telescopes (IACTs)~\citep{hess_nam,Backes_multi_wave}. Although the Gamsberg Mountain has no major observatory yet, the meteorological conditions were previously established to be suitable for astronomy\citep{eso_1994_5, GNSS_lott}.\\
\\
Precipitable water vapour~(PWV) is the main source of opacity at millimetre and submillimetre wavelengths as it absorbs emissions at these wavelengths~\citep{pwvdefinition}. Moreover, the presence of PWV causes phase delay on millimetre and submillimetre astronomical signals which may lead to a decline in performance on millimetre and submillimetre long baseline interferometers if not corrected~\citep{phase_delay_alma}. \citet{fransisco_thesis} conducted a study in which different instruments at the H.E.S.S. site were used to determine the PWV. The instruments in the study which included of the AErosol RObotic NETwork~(AERONET)~\citet{aeronet}, Autonomous
Tool for Measuring Observatory site COnditions PrEcisely~(Atmoscope)~\citep{Atmoscope}, and H.E.S.S. weather station and radiometers~\citep{hess_nam} all had their biases based on the conditions they take measurements. For example, the AERONET only records during the day during cloudless periods and the CT radiometers only record during during photometric nights which results in the PWV from these instruments being biased towards lower values. Furthermore, because of the different conditions and periods under which the instruments record measurements, they could not be properly validated against each other and, therefore, their results could not be verified.\\
\\
In order to assess the PWV and find the site with the best PWV conditions among the potential sites for the AMT, a Global Navigation Satellite System~(GNSS) station was installed at the H.E.S.S. site and the Gamsberg Mountain~\citep{GNSS_lott}. GNSS stations are known to be a viable and reliable alternative to the more conventional use of radiometers in measuring PWV. \citet{gnss_thesis_combrink} looked into verifying GNSS-derived PWV measurements against those from very-long baseline interferometry~(VLBI) observations of the 26-metre Hartebeesthoek radio telescope and a Water Vapour Radiometer~(WVR) 2000 at the Hartebeesthoek radio astronomy observatory~(HARTRAO). A positive correlation of above 90\% was found between the GNSS-derived PWV measurements and that determined by different methods, such as VLBI observations and WVR 2000~\citep{gnss_thesis_combrink}. Furthermore, a recent study conducted by~\citet{GNSS_atacama} using GNSS station data in the Atacama desert to measure the PWV for submillimetre and millimetre observations proved that GNSS station-derived PWV data could be reliably used for site evaluation and analysis. A comparison between the GNSS station-derived PWV data to scaled radiometer-derived PWV data found the instruments to have a mean offset of 0.64~mm over 15~minutes. Moreover, with the availability of the scaled PWV measurements from the radiometer, a weighted mean temperature~($\mathrm{T_m}$) model with respect to the surface temperature~($\mathrm{T_s}$) was derived for the Atacama. The $\mathrm{T_m}$ model allowed the local $\mathrm{T_s}$ to be incorporated in the calculation of the GNSS-derived PWV at the Atacama~\citep{GNSS_atacama}.\\
\\
The PWV results at the H.E.S.S. site and Gamsberg Mountain based on the GNSS station measurements by~\citet{GNSS_lott} found that both sites have the potential for millimetre wave astronomy with an overall median PWV of 14.27~mm and 9.25~mm, respectively. During the window of observation for the Event Horizon Telescope~(EHT) which typically occurs from March through April, a median PWV value of 16.62~mm and 11.20~mm was determined at the H.E.S.S. site and the Gamsberg Mountain, respectively. The study concluded based on the GNSS PWV data that Gamsberg Mountain had the lowest PWV conditions among the two potential sites. As there was no available site or regional model for $\mathrm{T_m}$, the $\mathrm{T_m}$ values used to calculate the PWV of GNSS stations at the H.E.S.S. site and the Gamsberg Mountain were rather interpolated from the European Centre for Medium-Range Weather Forecasts~(ECMWF) by the Nevada Geodetic Laboratory~(NGL) as opposed to being determined locally on site~\citep{GNSS_nevada,europe_weather_forecast}.\\
\\
According to~\citet{GNSS_lott}, MERRA-2 data consist of upper air measurements in addition to satellite and surface measurements assimilated into earth system modelling. The study by~\citet{GNSS_lott} validated MERRA-2 PWV against GNSS station PWV measurements at the H.E.S.S. site and found a 92\% correlation and a percentage difference of 7.45\% between the two measurements. The difference was attributed to the fact that the MERRA-2 PWV measurements are mainly based on satellite data and also involve interpolation in processing the data to the H.E.S.S. location, which therefore does not take well into account the local on-site conditions~\citep{GNSS_lott}. Studies by~\citet{Potential_sites_Ray} and ~\citet{astr_astro_site_study} also looking into potential sites for the next generation~(ng) EHT using MERRA-2 data also reported similar discrepancy between PWV from MERRA-2 data and that obtained insitu on some of the potential sites. For example, ~\citet{astr_astro_site_study} attributed the offset difference at Cerro Paranal in his study to strong local wind conditions due to it being close to the Pacific Ocean.\\
\\
The above reasons highlight the need of additionally using an instrument such as a radiometer on site as a way of validating any PWV results from a test instrument as oppose to using MERRA-2 data alone. In this study, a 210~GHz WVR was installed insitu at the H.E.S.S. site and used to investigate the accuracy and reliability of the GNSS-derived PWV results by~\citet{GNSS_lott} of both the H.E.S.S. site and the Gamsberg Mountain. Since the $\mathrm{T_m}$ used in the calculation of the GNSS PWV was determined through interpolation by the NGL, the study would also look into improving the GNSS station PWV measurement by developing a $\mathrm{T_m}$ with respect to on-site $\mathrm{T_s}$ for the H.E.S.S. site and the region in general by using the 210~GHz WVR and GNSS station data.

\section{Methods}
\begin{figure}
    \centering
    \includegraphics[scale = 0.17, trim=0 8cm 7cm 0, clip]{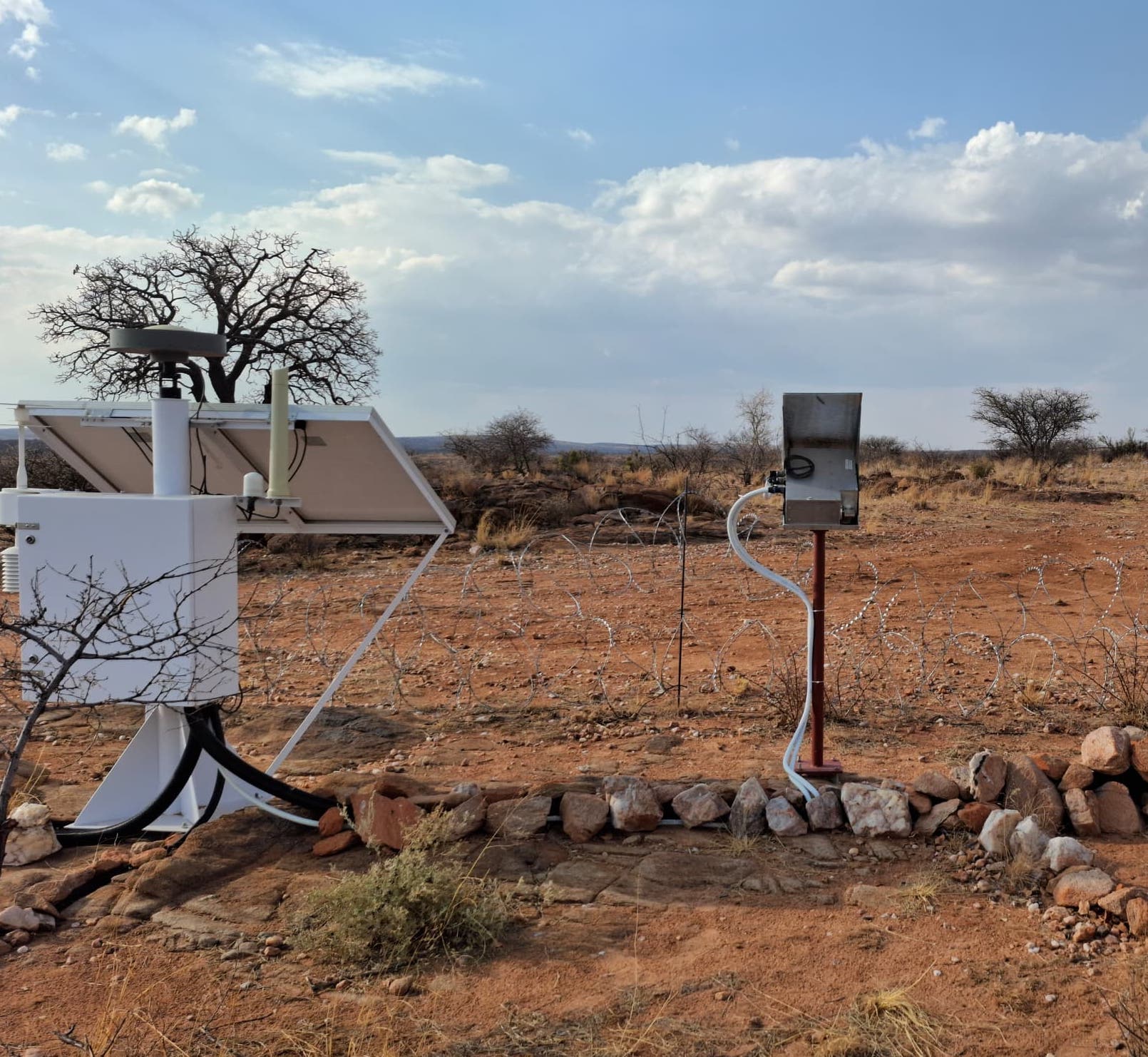}
    \caption{GNSS station (\emph{left}, in white) installed in 2022 and 210~GHz WVR (\emph{right}) installed in 2024 to validate GNSS PWV data at the H.E.S.S. site.}
    \label{fig:radiometer_gnss}
\end{figure}

Since both GNSS stations installed at Gamsberg Mountain and at the H.E.S.S. site are identical in design and the same methods were applied in calculating the PWV from the GNSS stations from both sites, only the PWV from the GNSS station installed at the H.E.S.S. site will be validated. The results of the validation will also apply to the PWV calculated from the GNSS station at the Gamsberg Mountain. The GNSS station at the H.E.S.S. site was installed in September 2022 and has consistently taken data. A 210~GHz WVR that measures opacity was installed in March~2024 alongside the GNSS station~(as depicted in Figure~\ref{fig:radiometer_gnss}) in order to validate the PWV measurements from the GNSS station. 
\\
\\
This section will describe the methods applied to the raw data in order to have PWV as a product of both the GNSS station and 210~GHz WVR.
\subsection{MERRA-2}
\begin{figure}%
    \centering
    \subfloat[\centering The red represents PWV vs opacity at 210~GHz whilst the blue line represents the fit of the form in equation~\ref{eq:poly_fit} on the data. This model will be used to convert 210~GHz WVR opacity into PWV.]{{\includegraphics[scale=0.3]{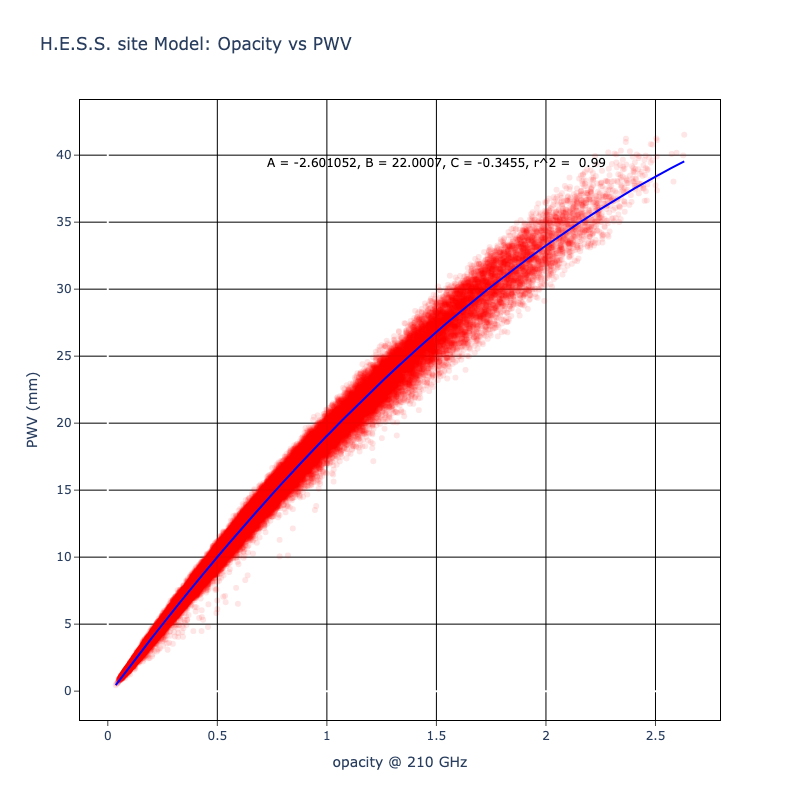} }
    \label{fig:model_pwv}}%
    \qquad
    \subfloat[\centering The blue represents PWV vs opacity at 210~GHz whilst the red line represents the fit of the form in equation~\ref{eq:poly_fit} on the data. This model will be used to convert GNSS PWV to opacity at 210~GHz.]{{\includegraphics[scale=0.3]{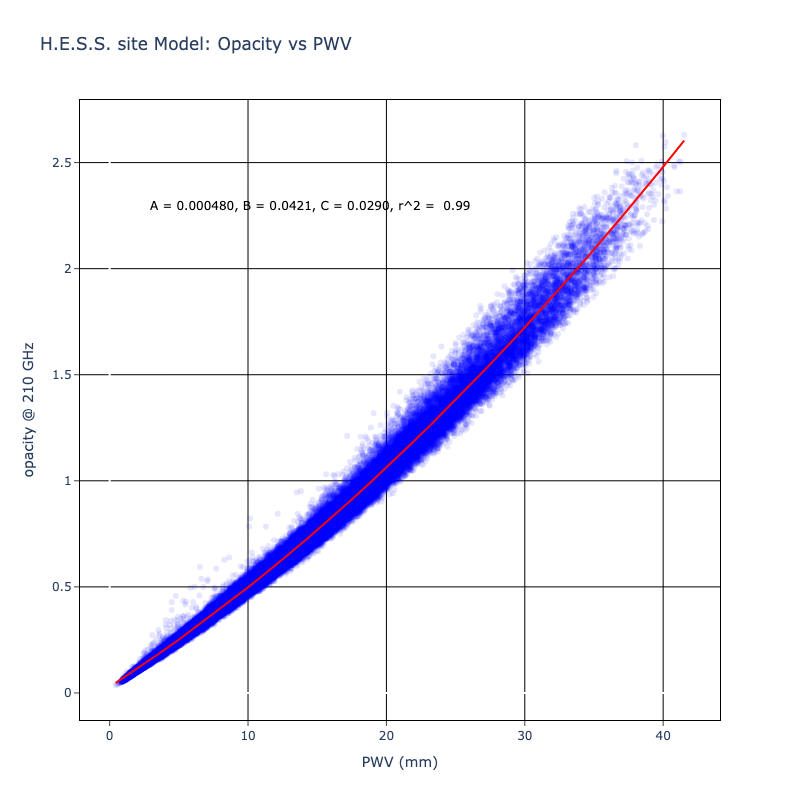} }
    \label{fig:model_opacity}
    }%
    \caption{Models between PWV vs opacity at 210~GHz based on 24 years of MERRA-2 data at the H.E.S.S. site.}%
    \label{fig:PWV_opacity_models}%
\end{figure}

Since the 210~GHz radiometer measures the optical depth/opacity $\tau_{0}$ at zenith at a frequency of 210~GHz, this opacity had to be converted to PWV in order to properly compare with the GNSS PWV data and equally vice versa. To do this, a model of PWV against opacity at 210~GHz was needed at the H.E.S.S. site. For this, MERRA-2 data were used, with the processes applied to acquire the MERRA-2 data at the H.E.S.S. site previously described by~\citet{GNSS_lott}. In this study, the MERRA-2 dataset ranges over a 24-years period between 2000 and 2024 with a temporal resolution of 3~hours. The atmospheric models of PWV against opacity at 210~GHz were then created using the 24 years of MERRA-2 data as shown in Figure~\ref{fig:PWV_opacity_models}.
A polynomial fit of the form,
\begin{equation}
y = Ax^{2} + Bx + C
\label{eq:poly_fit}
\end{equation}
was fitted with the coefficients provided in Table~\ref{tab:PWV_opacity210_hess}. The coefficients of determination~$R^2$, is the measure of how well the model can statistically predict the outcome. Both models for either converting opacity to PWV or visa versa have coefficients of determination of 0.99. This means that the model fits for 99\% of the data.
\begin{table}
\caption{Coefficients of the polynomial in equation~\ref{eq:poly_fit} when fitted to PWV against opacity~($\tau_{0}$) at 210~GHz at the H.E.S.S. site and its coefficient of determination~$R^2$.}
\label{tab:PWV_opacity210_hess}
 \centering
 \begin{tabular}{c c c c c c c}
		\hline
		$y$ & $x$ & $A $ & $ B $ & $C$ & $r^2$\\ [0.5ex] 
		\hline
		PWV & $\tau_{0}$ & -2.601052~mm &  22.0007~mm &  -0.3455~mm & 0.99\\  
		$\tau_{0}$ & PWV & 0.000480~mm$^{-2}$ &  0.0421~mm$^{-1}$ &  0.029 & 0.99\\
		\hline
	\end{tabular}
\end{table}

\subsection{GNSS station}
\begin{figure*}
    \centering
    \includegraphics[scale = 0.7]{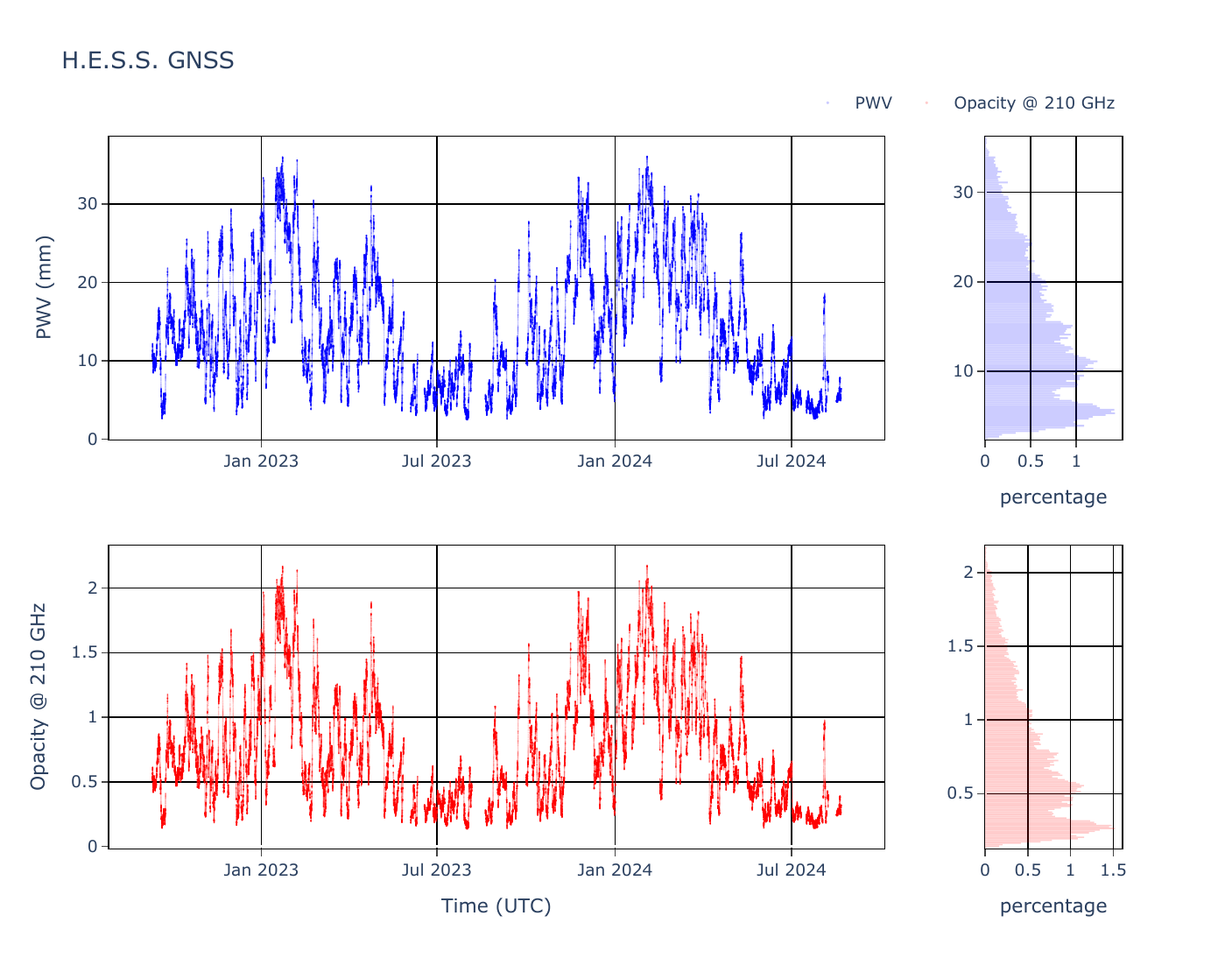}
    \caption{GNSS station PWV and opacity as deduced from the relationship in Figure~\ref{fig:model_opacity} at the H.E.S.S. site. The NGL $\mathrm{T_m}$ and on site pressure was used in the calculation of PWV.}
    \label{fig:HESS_opacity_and_PWV_timeseries_GNSS}
\end{figure*}

The details of the methods used in this study to retrieve PWV data from GNSS station measurements were previously described by~\citet{GNSS_lott}. According to~\citet{GNSS_lott}, the PWV data were calculated using the Zenith Total Delay~(ZTD)
and the $\mathrm{T_m}$ provided by the NGL~\citet{GNSS_nevada} while the pressure used was measured by the MET 4A weather station, which is integrated within the GNSS station. The ZTD is the total delay of signal from satellite at zenith position to the GNSS station receiver and is calculated as the sum of the Zenith Hydrostatic delay~(ZHD) and the Zenith Wet Delay~(ZWD),
\begin{equation}
\mathrm{ZTD = ZHD + ZWD}
\label{eq:ZTD}
\end{equation}
where the ZHD is then given by,
\begin{equation}
\mathrm{ZHD} = (2.2779\pm0.0024)\times\frac{P_s}{f(\lambda,H_s)}
\label{eq:ZHD}
\end{equation}
where $P_s$ is the on site surface pressure in hPa and $f(\lambda,H_s)$ is defined as,
\begin{equation}
f(\lambda,H_s) = 1 - 0.00266\times\cos({2\lambda})-0.00028H_s
\label{eq:function}
\end{equation}
where $H_s$ is the height in metres and~$\lambda$ the latitude of the GNSS station. Given the NGL ZTD products and the calculated ZHD using the pressure~$P_s$ measured on site by the MET4A weather station, the ZWD could then be calculated as,
\begin{equation}
\mathrm{ZWD = ZTD - ZHD}
\label{eq:ZWD}
\end{equation}
With the NGL~$\mathrm{T_m}$, the PWV was then determined from the ZWD with equation~\ref{eq:H_const} and ~\ref{eq:PWV_zwd} as described under subsection~\ref{sec:PWV}. The raw data measurements of the GNSS station are integrated over 30-second intervals but were resampled to 5~minutes intervals by NGL. Figure~\ref{fig:HESS_opacity_and_PWV_timeseries_GNSS} shows the time series PWV data from the GNSS station at the H.E.S.S. site.
Using the coefficients of the relationship derived in Figure~\ref{fig:model_opacity} tabulated in Table~\ref{tab:PWV_opacity210_hess} and the relation~\ref{eq:poly_fit}, the PWV was converted in opacity at 210~GHz.

\subsection{210~GHz Water Vapour Radiometer}
\begin{figure*}
    \centering
    \includegraphics[scale = 0.7]{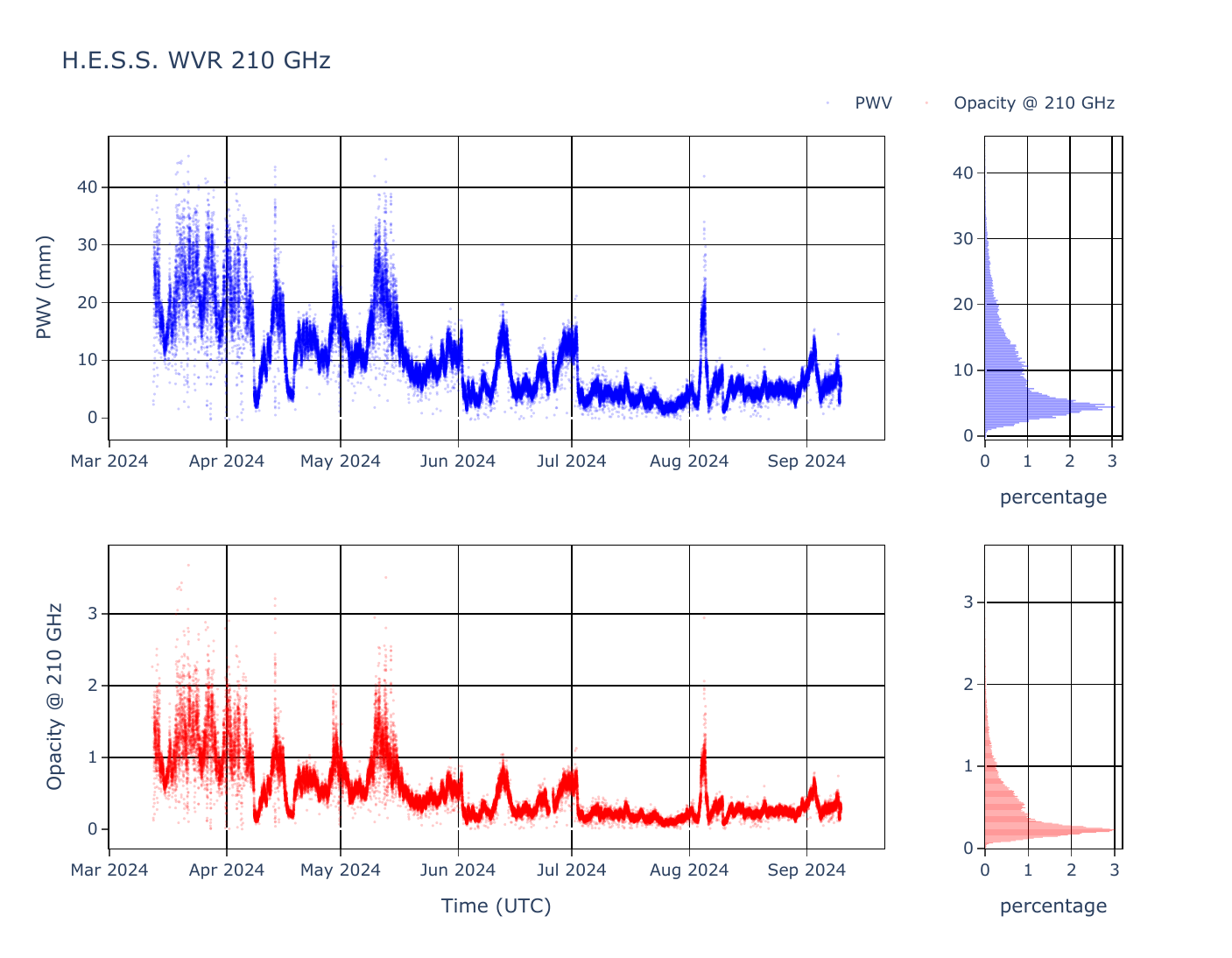}
    \caption{210~GHz Radiometer opacity and PWV as deduced from Figure~\ref{fig:model_pwv} at the H.E.S.S. site.}
    \label{fig:HESS_opacity_and_PWV_timeseries210}
\end{figure*}

The 210~GHz WVR takes opacity measurements with an integration time that ranges between one measurement per~5 to~15 minutes. According to~\citet{Hiriate_215_Radiometer}, in the absence of the cosmic background, the antenna temperature $T_\text{ant}$ of the 210~GHz WVR is given by,
\begin{equation}
T_\text{ant} = \eta T_\text{atm}\left(1-e^{-\tau}\right),
\label{eq:antenna_temp}
\end{equation}
where~$\eta$ is the coupling factor between the 210~GHz WVR and the antenna, $T_\text{atm}$ is the atmospheric temperature and~$\tau$ is the opacity. The total sky temperature~$T_\text{sky}$ of the system is given as the sum of antenna temperature~$T_\text{ant}$ and the receiver noise temperature $T_\text{rec}$, then the antenna temperature~$T_\text{ant}$ can then be written as
\begin{equation}
T_\text{ant} = T_\text{sky} - T_\text{rec}
\label{eq:antenna_temp_system}
\end{equation}
inserting equation~\ref{eq:antenna_temp} into equation~\ref{eq:antenna_temp_system}, we can rewrite equation~\ref{eq:antenna_temp_system} as,
\begin{equation}
T_\text{sky} = \eta T_\text{atm}\left(1-e^{-\tau}\right) + T_\text{rec}.
\label{eq:antenna_temp_re-written}
\end{equation}
The temperature load reference is given by,
\begin{equation}
T_\text{ref} = \eta T_\text{L} + T_\text{rec}
\label{eq:load_ref}
\end{equation}
solving for the receiver temperature $T_\text{rec}$ in equation~\ref{eq:load_ref} and inserting it into equation~\ref{eq:antenna_temp_re-written}, the expression becomes,
\begin{equation}
T_\text{ref} - T_\text{sky} = \eta \rho \left( T_\text{L} - T_\text{atm}\left(1-e^{-\tau}\right)\right),
\label{eq:ref_sky}
\end{equation}
where~$\rho$ is the factor relating the voltage output of the amplifier to the antenna temperature. The comparison between temperature and voltage yields,
\begin{equation}
V_\text{ref} - V_\text{sky} = \eta \rho \left( T_\text{L} - T_\text{atm}\left(1-e^{-\tau}\right)\right).
\label{eq:ref_sky_voltages}
\end{equation}
Under an isothermal atmosphere, the ambient temperature around the receiver will be the same everywhere, thus $T_\text{Load} \approx T_\text{atm}$, this reduces the expression to,
\begin{equation}
V_\text{ref} - V_\text{sky} = \eta \rho T_\text{atm}e^{-\tau}.
\label{eq:ref_sky_voltages_short}
\end{equation}
The optical depth in any direction of the sky can be expressed in terms of the optical depth as zenith $\tau_\text{0}$ as,
\begin{equation}
\tau = \tau_\text{0}\sec\left(z\right),
\label{eq:optical_depth}
\end{equation}
where~$z$ is the zenith angle. Inserting equation~\ref{eq:optical_depth} into equation~\ref{eq:ref_sky_voltages_short} and calculating the natural log on both sides of the equation yields,
\begin{equation}
\ln\left(V_\text{ref} - V_\text{sky}\right) = -\tau_\text{0}\sec\left(z\right) + \ln\left(\eta \rho T_\text{atm}\right).
\label{eq:voltages_temp_opacity}
\end{equation}
For an isothermal atmosphere, there is a linear relationship between the logarithm of the differences of voltages and the zenith distance and can also be shown as,
\begin{equation}
\ln\left(\frac{V_\text{ref} - V_\text{sky}}{\eta \rho T_\text{atm}}\right) = -\tau_\text{0}\sec\left(z\right).
\label{eq:voltages_temp_opacity_linear}
\end{equation}
Therefore, the slope of the linear fit can be determined as the optical depth~$\tau_{0}$ at zenith. Although this method described by ~\cite{Hiriate_215_Radiometer} was for the 215~GHz WVR, the 210~GHz WVR in this study uses the same method. Figure~\ref{fig:HESS_opacity_and_PWV_timeseries210} shows the time series of opacity taken by the 210~GHz WVR and the converted PWV data using the fit in Figure~\ref{fig:model_pwv}, coefficient in Table~\ref{tab:PWV_opacity210_hess} and equation~\ref{eq:poly_fit}.
It is evident from Figure~\ref{fig:HESS_opacity_and_PWV_timeseries210} that there appears to be scatter in the 210~GHz WVR data which seems unphysical.

\section{Results}
\subsection{GNSS and 210~GHz Radiometer Comparison}
\begin{figure*}
    \centering
    \includegraphics[scale = 0.6]{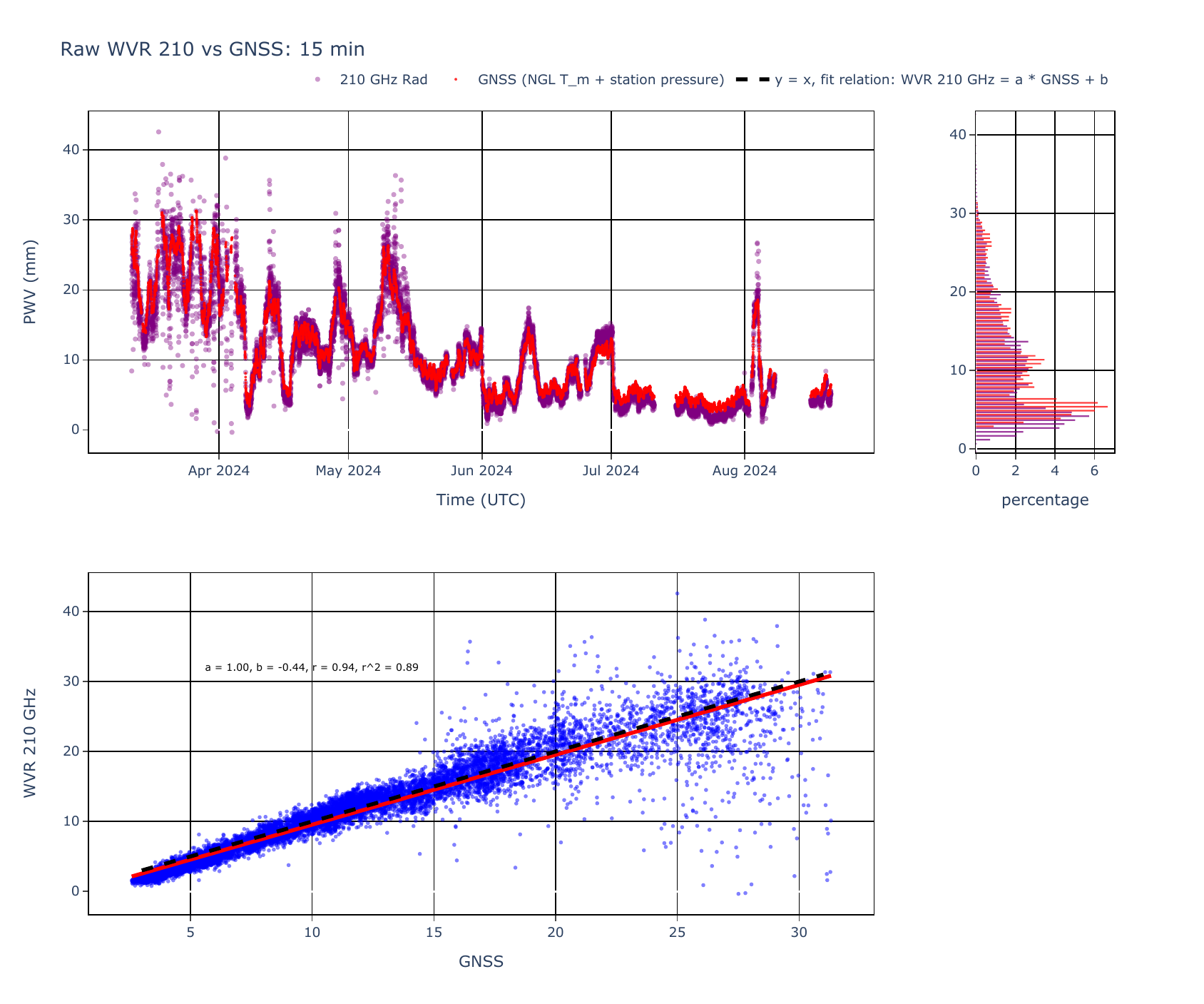}
    \caption{210~GHz Radiometer and GNSS station PWV measured over the same period at the H.E.S.S. site.}
    \label{fig:PWV_WVR}
\end{figure*}

\begin{figure*}%
    \centering
    \subfloat[\centering Gaussian fitted to the difference in PWV between the 210~GHz WVR and GNSS sation. Values having a difference less than -3.52~mm and more than 4.21~mm was flagged.]{{\includegraphics[scale=0.5]{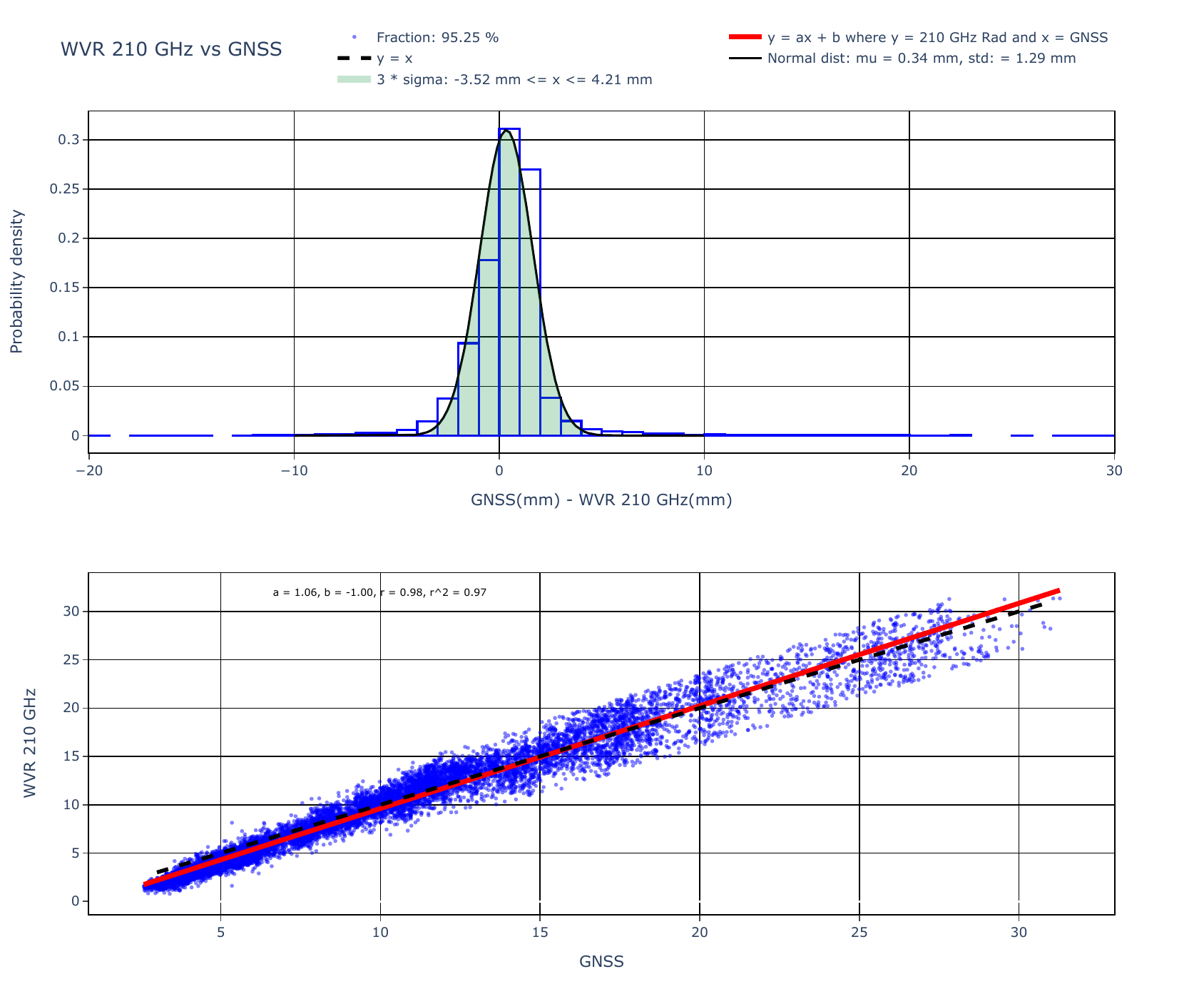} }
    \label{fig:PWV_WVR_clean}}%
    \qquad
    \subfloat[\centering Resultant PWV of GNSS station and 210~GHz after flagging.]{{\includegraphics[scale=0.55]{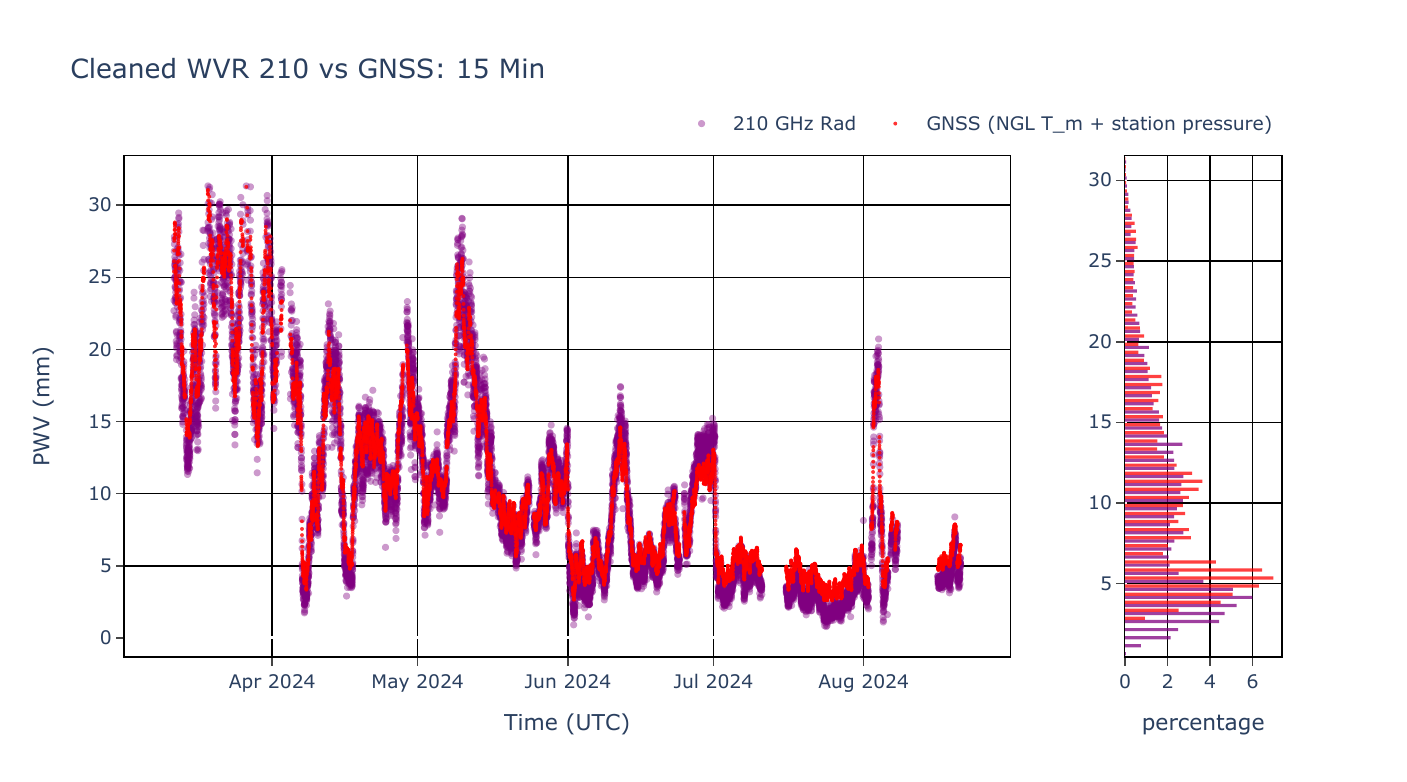} }
    \label{fig:PWV_WVR_cleaned_data}
    }%
    \caption{GNSS station and the 210~GHz WVR data after flagging data were the difference of the measurements between the GNSS station and 210~GHz is outside $3\sigma$.}%
    \label{fig:GNSS_WVR_ngl}%
\end{figure*}

In order to compare the GNSS station and 210~GHz WVR data, both datasets were converted into 15~minutes integration period. The overlapping periods for which data were measured by both instruments were then determined as can be seen in Figure~\ref{fig:PWV_WVR}. In this study, even though the opacity at 210~GHz can be used to assess the measurements of both instruments, only the PWV was used to determine the accuracy of the two instruments in-line with the evaluation of the PWV results of the study by~\citet{GNSS_lott}.
It is quite evident from Figure~\ref{fig:PWV_WVR} that the GNSS station and the 210~GHz WVR have the same trend in time with the radiometer data seemingly marginally higher. The scatter in the 210~GHz WVR is even more evident as they do not appear in the GNSS station data which is an indication that these points aren't physical but possibly an instrument systematic error. In order to find and remove these scattering data points that account for about 4.75\% of the data, a normal distribution of the difference in PWV between the GNSS station and the 210~GHz WVR was found as can be seen in Figure~\ref{fig:PWV_WVR_clean}.
Simultaneous PWV measurements from both instruments that resulted in a PWV difference greater than $3\sigma$, which is less than -3.52~mm and that greater than 4.21~mm were flagged. This resulted in 95.25\% of the data being retained. In doing this, the scatter points in the 210~GHz WVR was removed as can be seen in Figure~\ref{fig:PWV_WVR_clean}. A correlation of 98\% was found with a mean offset of 0.35~mm between the instruments. It is quite evident from the fit plot that the two instruments are almost one to one. The resulting data after flagging are shown in Figure~\ref{fig:PWV_WVR_cleaned_data}, which shows a similar trend in time between the instruments.

\subsection{Improving results with on site $\mathbf{T_m}$}
\label{sec:PWV}
\begin{figure*}
    \centering
    \includegraphics[scale = 0.5]{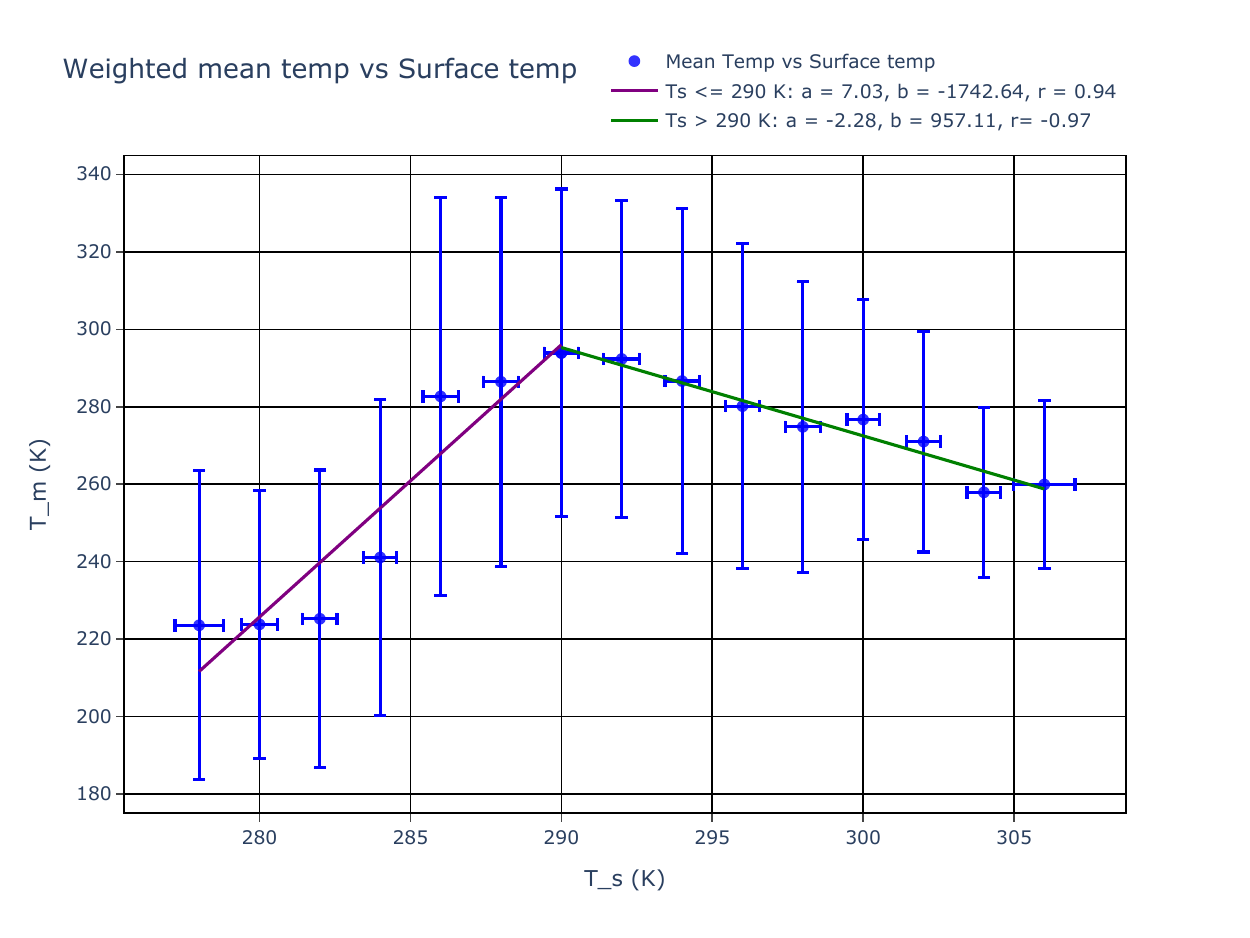}
    \caption{Relationship between $\mathrm{T_s}$ and $\mathrm{T_m}$ at the H.E.S.S. site.}
   \label{fig:Tm_plot}
\end{figure*}

The PWV results of the GNSS station can be improved by taking into account the $\mathrm{T_m}$ determined locally through on-site measurements rather than through interpolation as done by the NGL. PWV is given by,
\begin{equation}
\mathrm{PWV} = H * \mathrm{ZWD}
\label{eq:PWV_zwd}
\end{equation}
where $\mathrm{ZWD}$ is calculated from equation~\ref{eq:ZTD} and $H$ given by,
\begin{equation}
H\left(\mathrm{T_m}\right) = \frac{10^6}{\rho_w R_w \left( k^{'}_2 + k_3 \mathrm{T_{m}}^{-1} \right) }
\label{eq:H_const}
\end{equation}
where $\rho_w$ is the density of water given as 1000~kg\,m$^{-3}$, $R_w$ is the specific gas constant of water vapour, given as $461.4$ J\,K$^{-1}$\,kg$^{-1}$ and constants $k^{'}_2 = 22.1$~K\,hPa$^{-1}$ and $k_3 = 373900$~K$^2$\,hPa$^{-1}$~\citep{GNSS_atacama, pwv_zwd}. The $\mathrm{T_m}$ profile of a vertical column of water vapour is conventionally determined with radiosonde measurements and is given by,
\begin{equation}
\mathrm{T_m} = \frac{\int_{}^{} \frac{P_{w}}{T} \,dz}{\int_{}^{} \frac{P_{w}}{T^{2}} \,dz}
\label{eq:tm}
\end{equation}
$P_{w}$ and $T$ are the water vapour pressure and the temperature respectively~\citep{gnss_thesis_combrink,GNSS_atacama}. Due to the significant effect of the dry bias on the relative humidity measurements in arid regions, it becomes very challenging to measure the $\mathrm{T_m}$ using radiosonde~\citep{GNSS_atacama, radiosonde_dry}. The $\mathrm{T_m}$ can also be linearly related to the measured $\mathrm{T_s}$ in which it has the form of,
\begin{equation}
\mathrm{T_m} = a\mathrm{T_s} + b
\label{eq:linear_eq}
\end{equation}
as demonstrated by~\citet{GNSS_atacama}. Several regional models that relate the $\mathrm{T_s}$ to the $\mathrm{T_m}$ have been deduced before, such as the~\citet{bevis1992_T_m} for the United A
States~(US), \citet{egypt_tm} for Egypt~(EG), and the model by~\citet{GNSS_atacama} for the Atacama. This provides an alternative method for determining the $\mathrm{T_m}$ from the ground. The $\mathrm{T_m}$ at the H.E.S.S. site was modelled using the methods described in~\citet{GNSS_atacama}. The 210~GHz WVR provided the PWV data, and the GNSS station provided the ZTD and $\mathrm{T_s}$ for the analysis such as the one described in~\citet{GNSS_atacama}. Figure~\ref{fig:zwd_pwv_fit} provides the fits between the ZWD and the PWV and Figure~\ref{fig:Tm_plot} shows the resulting relationship between the $\mathrm{T_m}$ and the $\mathrm{T_s}$.
As evident from Figure~\ref{fig:Tm_plot}, there were two phases in which there was a positive correlation between the $\mathrm{T_m}$ and the $\mathrm{T_s}$ till 290~K after which a negative correlation was observed. A piecewise model was then developed for $\mathrm{T_m}$ as follow,
\begin{equation}
\mathrm{T_m}\left(\mathrm{T_s}\right) = 
\left\{
    \begin{array}{lr}
        7.03\mathrm{T_s} - 1742.64, & \text{if } \mathrm{T_s} \leq 290 K\\
        -2.28\mathrm{T_s} + 957.11, & \text{if } \mathrm{T_s} > 290 K
    \end{array}
\right\} = \mathrm{T_m}
\label{eq:Tm_piece}
\end{equation}
Using the on site $\mathrm{T_s}$ measured by the MET4A weather station on the GNSS station and the model in equation~\ref{eq:Tm_piece}, the $\mathrm{T_m}$ was then calculated. This now on site $\mathrm{T_m}$ were then used to calculate $H(\mathrm{T_m})$ in equation~\ref{eq:H_const} and subsequently the PWV using equation~\ref{eq:PWV_zwd}. These GNSS station PWV data which are based on insitu pressure and temperature were then compared to the 210~GHz WVR PWV measurements as can be seen in Figure~\ref{fig:Tm_plot_results}.\\
\\
The use of insitu temperature in the calculation of the PWV slightly improves the results of the GNSS station, as the offset reduced to 0.15~mm compared to 0.34~mm of when the NGL interpolated mean temperature was used. Table~\ref{tab:GNSS_vs_WVR} shows the offsets and standard deviation of PWV between the 210~GHz WVR and the GNSS station of when the NGL $\mathrm{T_m}$ and when the insitu derived $\mathrm{T_m}$ were used.  
\begin{table}
\caption{Comparison between 210~GHz WVR and the different GNSS station method.}
\label{tab:GNSS_vs_WVR}
	\centering
\begin{tabular}{c c c c}
    \hline
     Method & std [mm] & offset [mm]\\ [0.5ex]
    \hline
     NGL $\mathrm{T_m}$, site $P_s$ & 1.29 & 0.34\\ 
    Site $\mathrm{T_m}$ \& $P_s$ & 1.24 & 0.15\\
    \hline
\end{tabular}
\end{table}
These offsets in Table~\ref{tab:GNSS_vs_WVR} are much lower when compared to 0.64~mm obtained by~\citet{GNSS_atacama} over 15 minutes. The 0.34~mm and 0.15~mm are essentially negligible and suggest the 210~GHz WVR and GNSS station PWV data are one-to-one. The $\mathrm{T_m}$ model developed at the H.E.S.S. site shall also apply to the Gamsberg Mountain, as both sites are in the same region and experience the same meteorological conditions.

\section{Conclusions}
In this study, GNSS station-derived PWV measurements were validated against 210~GHz WVR-derived PWV measurements at the H.E.S.S. site. The results of this study also apply to the GNSS station at the Gamsberg Mountain as both stations employ similar methods in calculating PWV and are located in the same vicinity with similar meteorological conditions. We have shown a way to convert PWV into opacity at 210~GHz and similarly convert opacity at 210~GHz into PWV by modelling the relationship between PWV and opacity at 210~GHz from MERRA-2 data at the H.E.S.S. site. Using these models, the GNSS station PWV against that from WVR~210~GHz was compared to each other and analysed. We have also managed to develop a $\mathrm{T_m}$ model with respect to $\mathrm{T_s}$ at the H.E.S.S. site and similarly for the region. A high correlation of 98\% was observed between the PWV from the 210~GHz WVR and the GNSS station, with both instruments having similar trends over time. The offset PWV from the 210~GHz WVR and the GNSS was 0.34~mm when on site pressure from the MET4A was used to calculate the ZHD and the NGL interpolated $\mathrm{T_m}$ was used to calculate the PWV. The mean PWV offset between the two instruments improved by reducing to 0.15~mm when the on site $\mathrm{T_s}$ was used to calculate the $\mathrm{T_m}$ using the model developed in this study, which therefore provides a highly accurate alternative to calculate the PWV with the $\mathrm{T_s}$ measured on site. In this study, we showed that the GNSS station can be used with high accuracy to determine the PWV at the H.E.S.S. site and in the region with minimum offset from traditionally utilised instruments such as the 210~GHz WVR. Moreover, we have managed to show that the PWV results based on GNSS station measurements by~\citet{GNSS_lott} is a reliable assessment of the PWV at the H.E.S.S. site and at the Gamsberg Mountain with a minimal offset of 0.34~mm.

\section*{Acknowledgements}
This work was and is supported by multiple institutions including the South African Radio Astronomy Observatory~(SARAO) through Dr~Roelf Botha and the National Autonomous University of Mexico~(UNAM) through Prof.~Stanley E. Kurtz and his collaborators in CONACyT-NRF Project 291778./. We also thank the Nevada Geodetic Laboratory~(NGL) for processing and presenting the GNSS tropospheric products used in this study. This work has been partially supported by the ERC Synergy Grant \emph{BlackHolistic} and H2020-INFRADEV-2016-1 project 730884 JUMPING-JIVE.

\section*{Data Availability}
The GNSS, 210~Radiometer and MERRA-2 dataset supporting this study is available from the authors upon reasonable request. Alternatively, GNSS station product data files can be found on the Nevada Geodetic Laboratory link (\url{http://geodesy.unr.edu/NGLStationPages/stations/GBGA.sta}).



\bibliographystyle{rasti}
\bibliography{references} 



\appendix
\section{Modelling the weighted mean temperature}
This appendix shows the procedures used to derive the $\mathrm{T_m}$ model given by equation~\ref{eq:Tm_piece} for the H.E.S.S. site and the region. It furthermore gives a comparison plot between the 210~GHz WVR PWV data and that obtained from the GNSS station based on the derived insitu $\mathrm{T_m}$.
\subsection{Calculating the $\mathbf{T_m}$}
Following the procedures by ~\citet{GNSS_atacama}, only measurements of PWV, ZWD, and $\mathrm{T_s}$ taken at the same time when the relative humidity was greater than 40$\%$ were considered during the determination of $H(\mathrm{T_m})$. The ZWD and PWV were partitioned according to $\mathrm{T_s}$ as can be seen in Figure~\ref{fig:zwd_pwv_fit}, with the slope~$m$ of the fit relating ZWD and PWV given by,
\begin{equation}
m = \frac{\mathrm{ZWD}}{\mathrm{PWV}}
\label{eq:m_slope}
\end{equation}
from equation~\ref{eq:PWV_zwd}, $H$ can then be calculated from $m$ and is related to $m$ as,
\begin{equation}
H = \frac{\mathrm{PWV}}{\mathrm{ZWD}} = \frac{1}{m}
\label{eq:H_div}
\end{equation}
with $H$ and solving for $\mathrm{T_m}$ in equation~\ref{eq:H_const}, the $\mathrm{T_m}$ was then calculated for the partitioned $\mathrm{T_s}$. The resultant $\mathrm{T_m}$ and $\mathrm{T_s}$ are shown in Figure~\ref{fig:Tm_plot}, from which the relation between $\mathrm{T_m}$ and $\mathrm{T_s}$ in equation~\ref{eq:Tm_piece} was derived.
\begin{figure*}
\includegraphics[width=.32\textwidth, height =.35\textwidth]{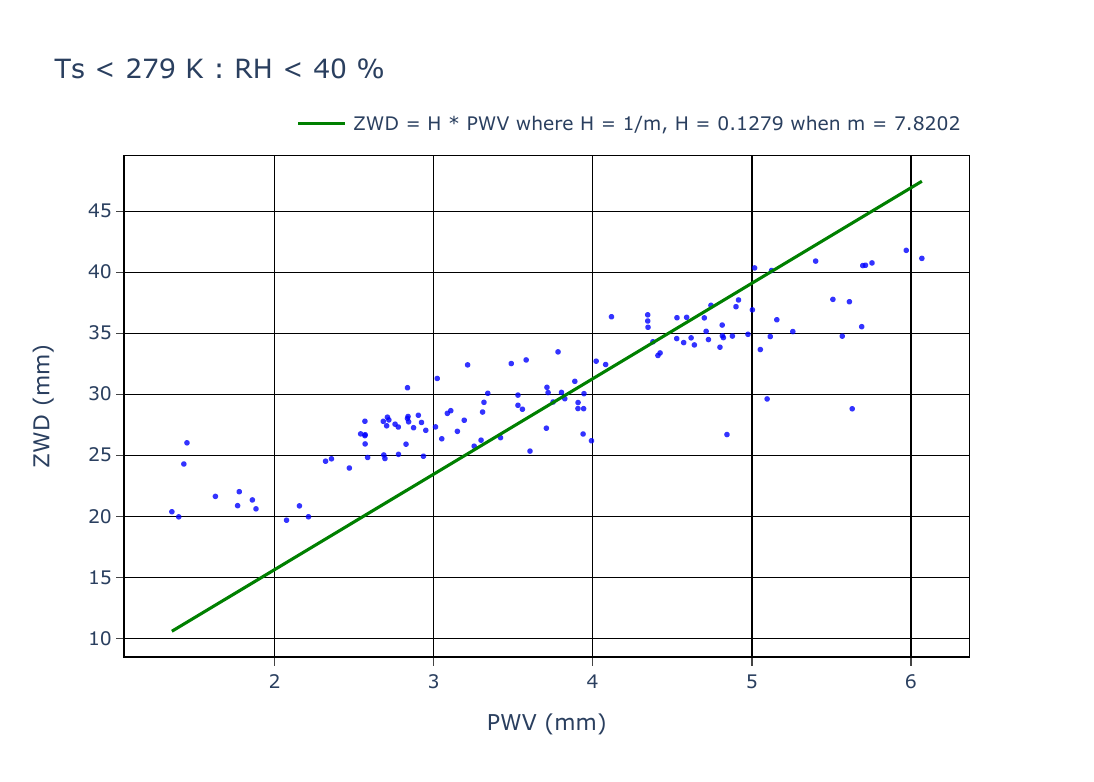}
\includegraphics[width=.32\textwidth, height =.35\textwidth]{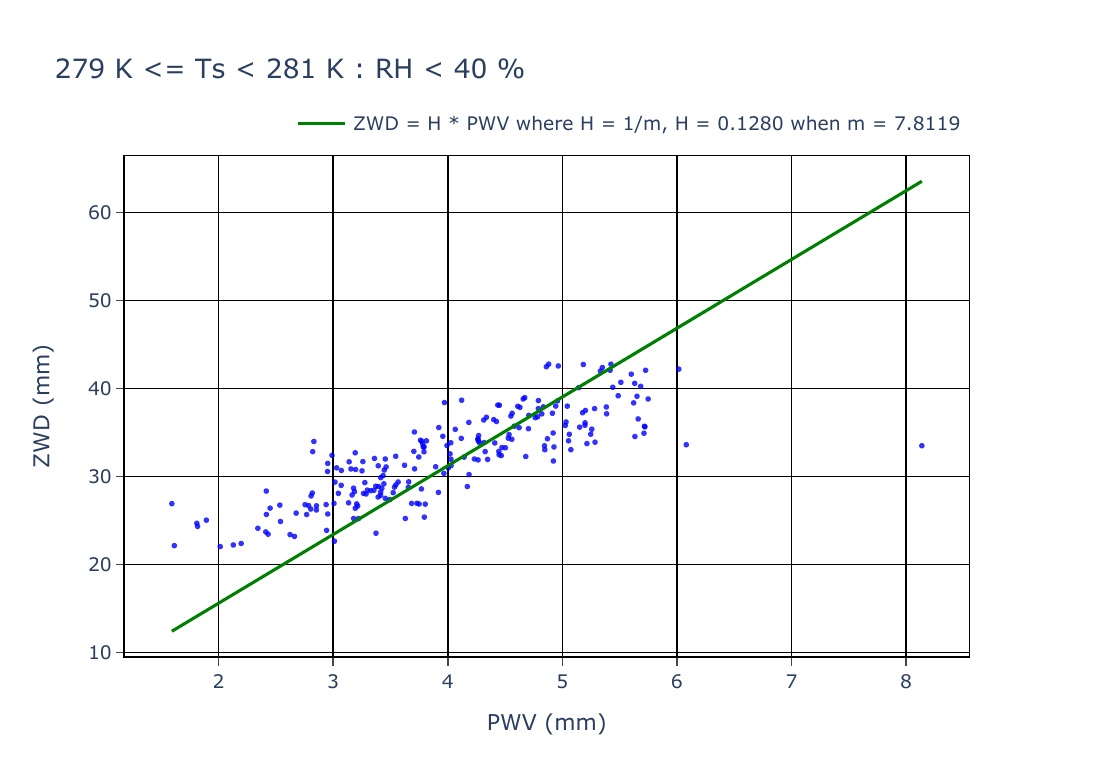}
\includegraphics[width=.32\textwidth, height =.35\textwidth]{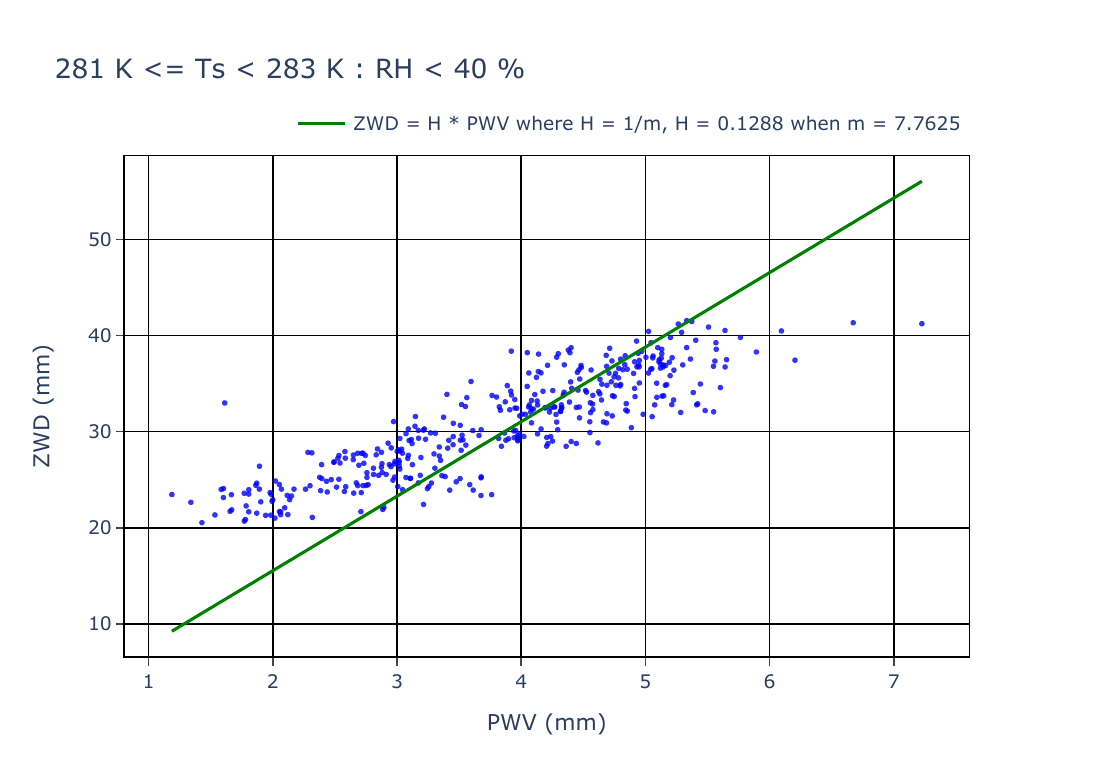}\quad

\medskip

\includegraphics[width=.32\textwidth, height =.35\textwidth]{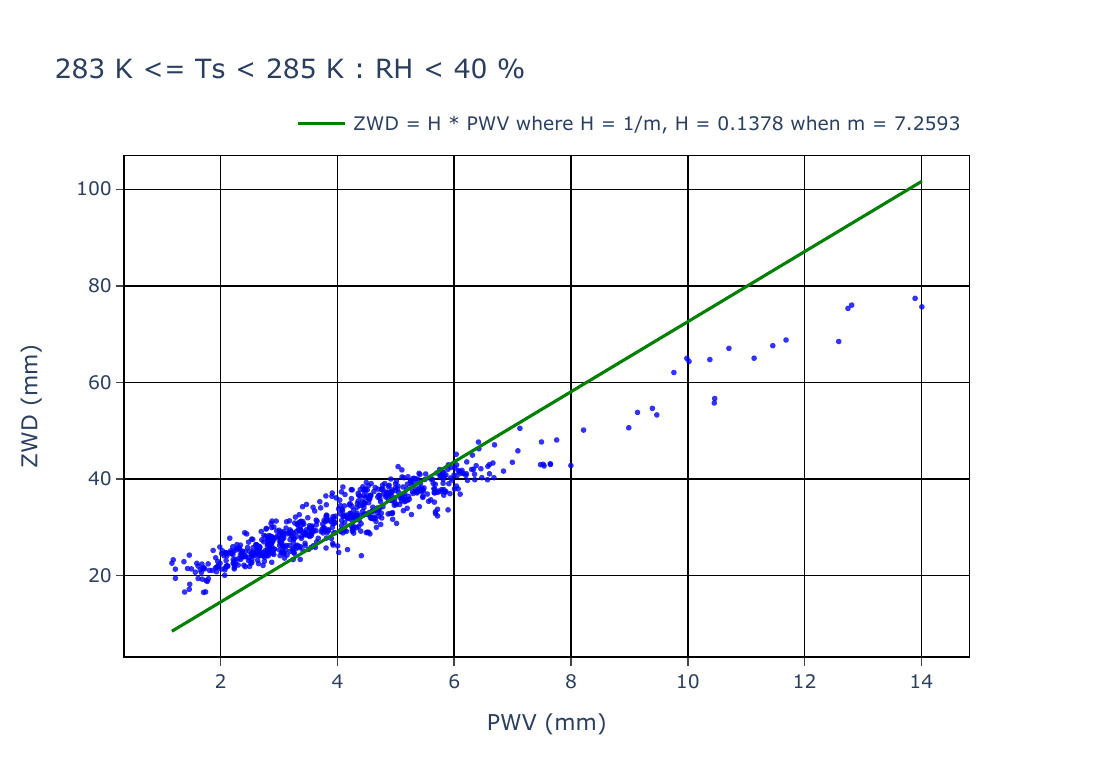}
\includegraphics[width=.32\textwidth, height =.35\textwidth]{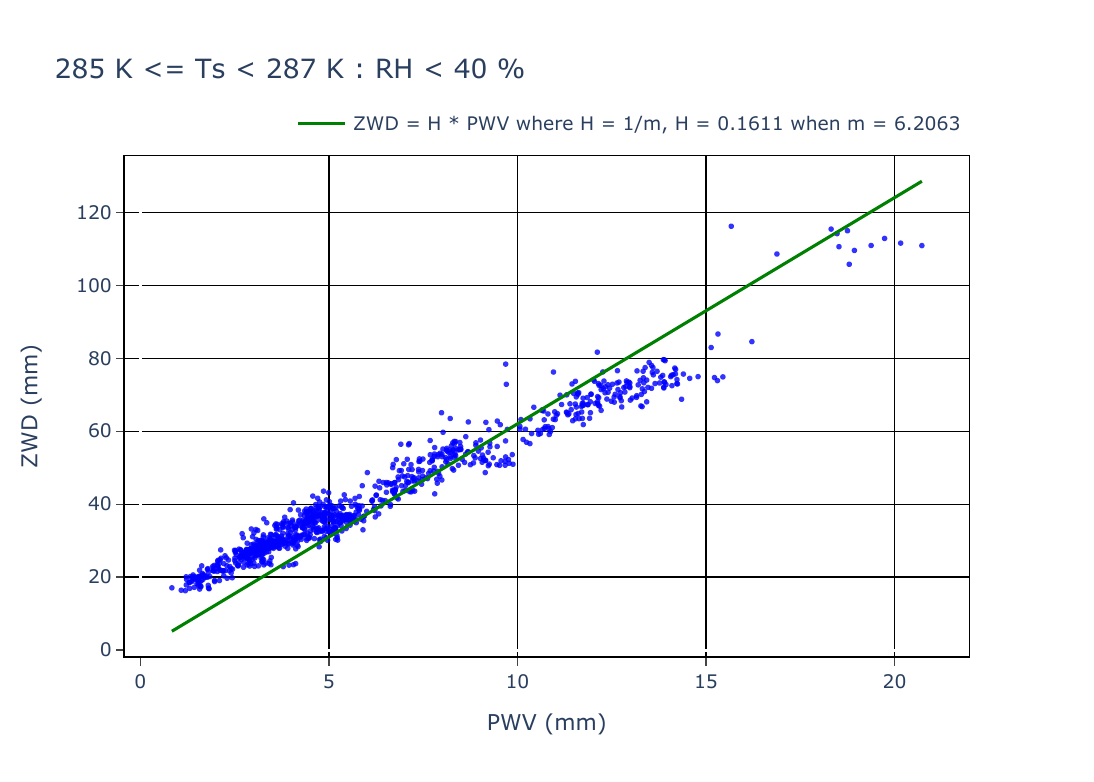}
\includegraphics[width=.32\textwidth, height =.35\textwidth]{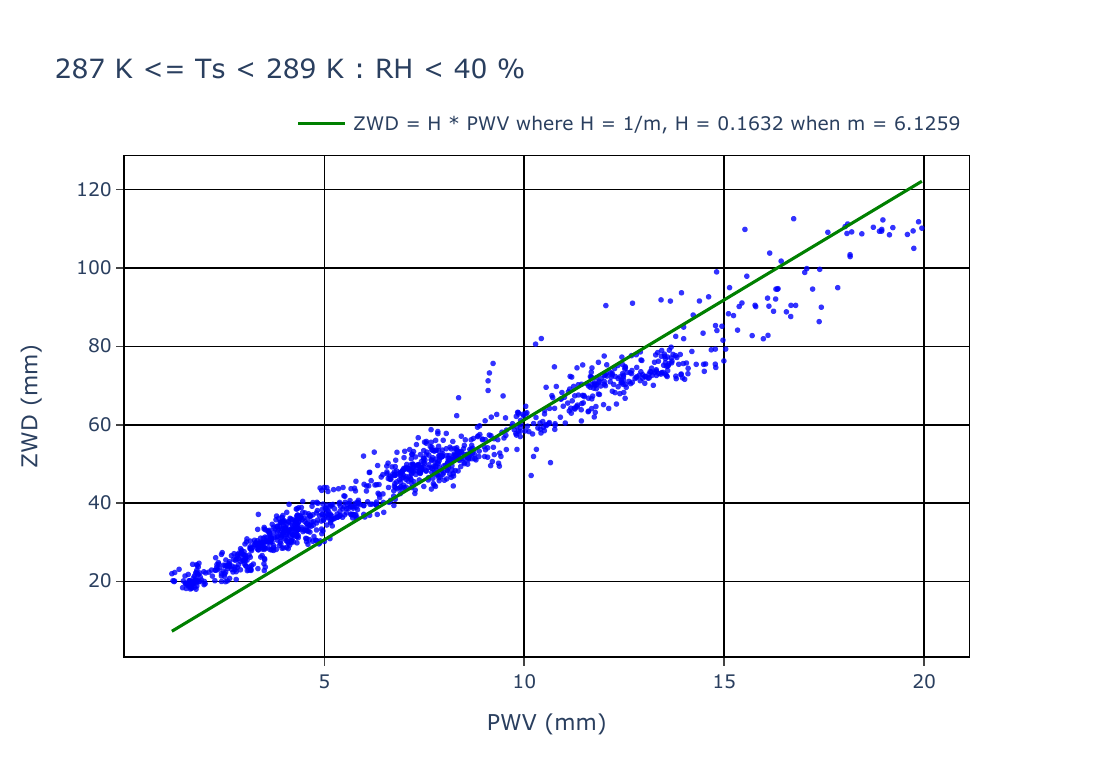}\quad

\medskip

\includegraphics[width=.32\textwidth, height =.35\textwidth]{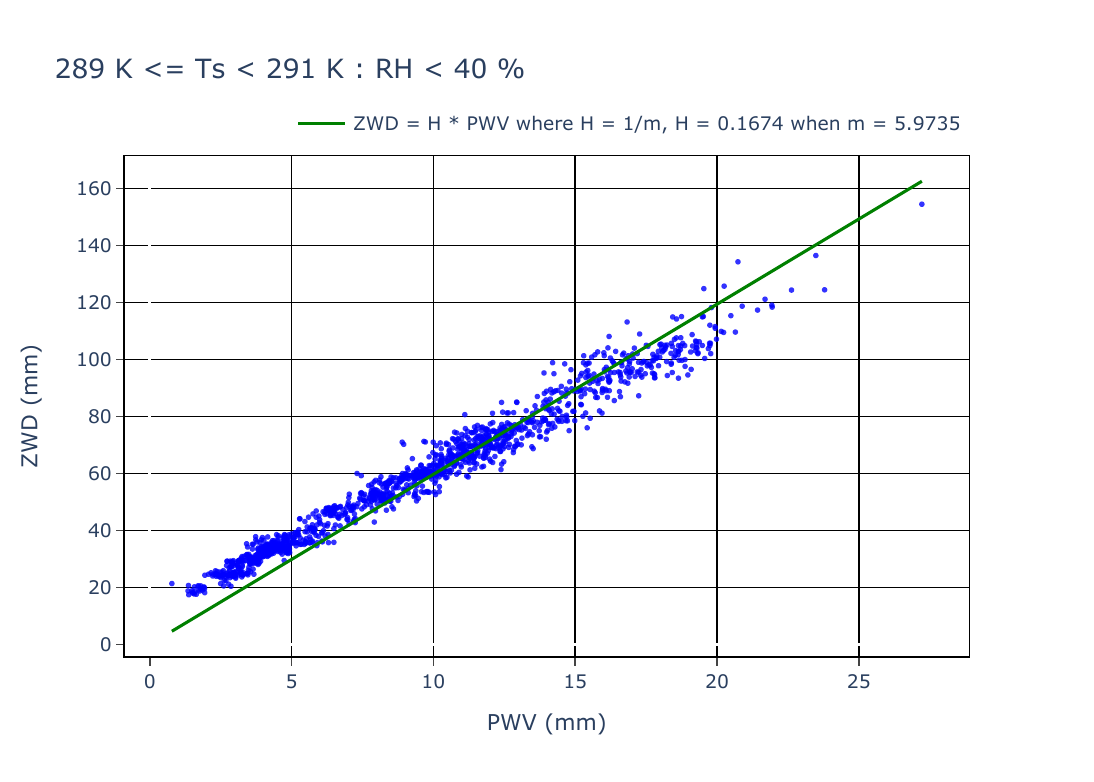}
\includegraphics[width=.32\textwidth, height =.35\textwidth]{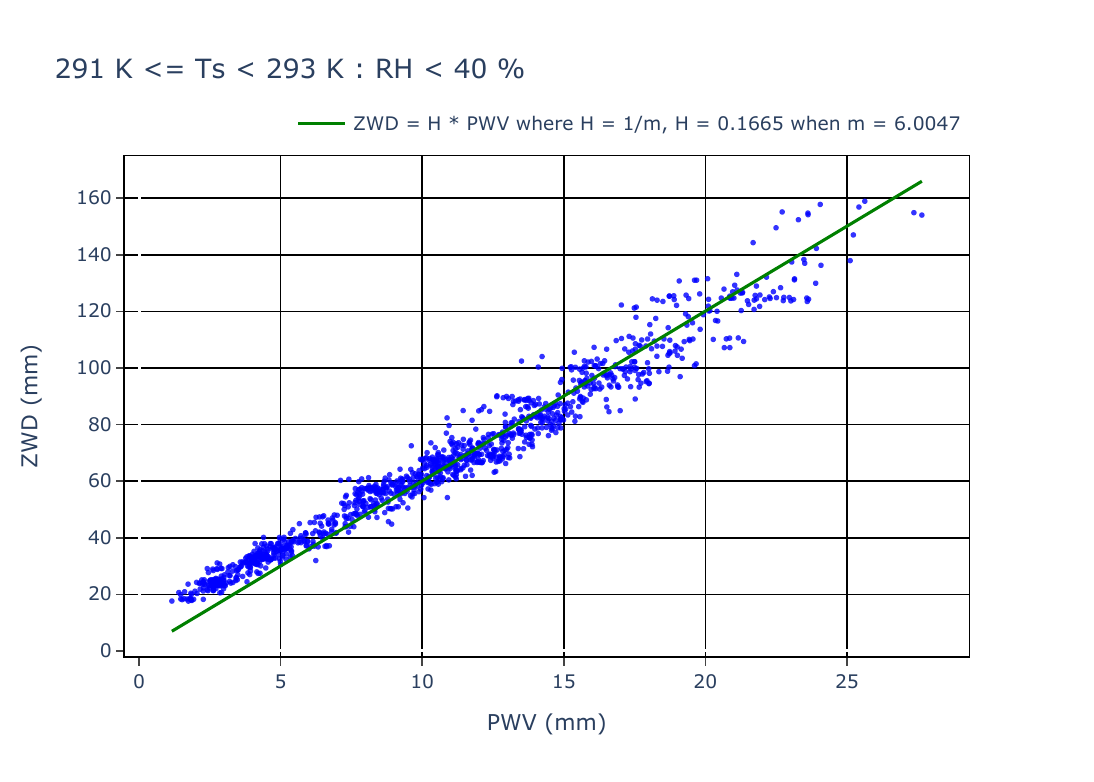}
\includegraphics[width=.32\textwidth, height =.35\textwidth]{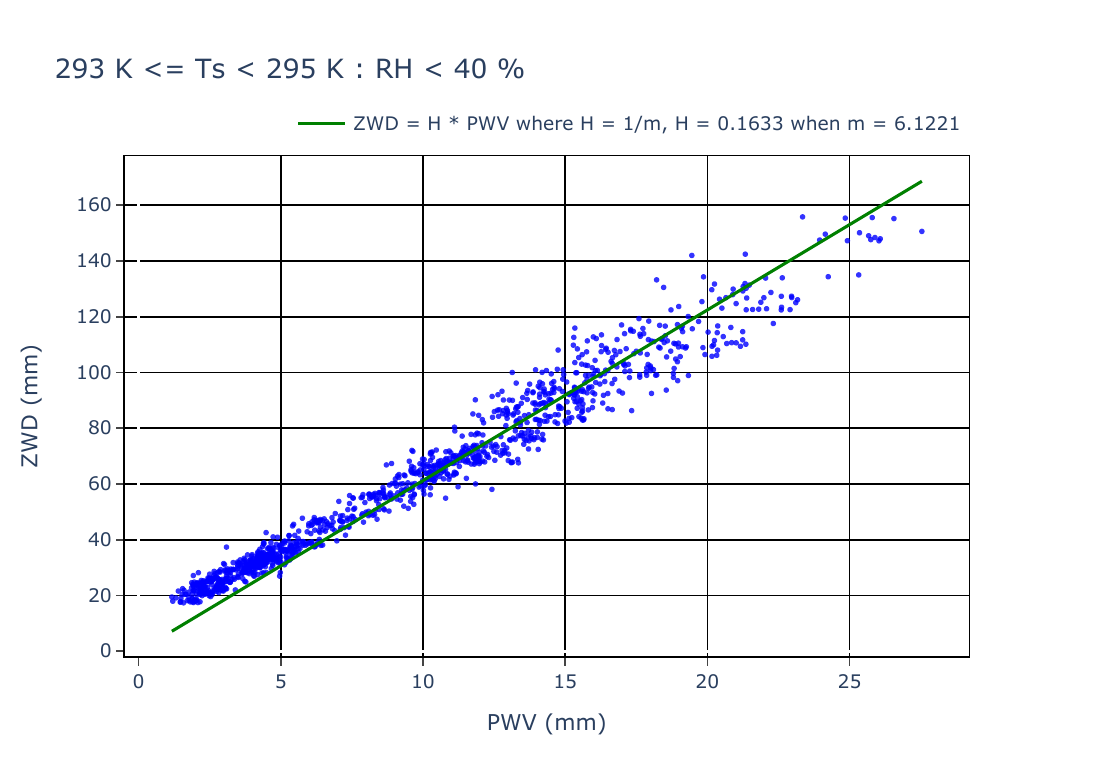}\quad
\caption{ZWD against PWV at different $\mathrm{T_s}$ at the H.E.S.S. site.}
\label{fig:zwd_pwv_fit}
\end{figure*}

\begin{figure*}
\ContinuedFloat 
\centering

\includegraphics[width=.32\textwidth, height =.35\textwidth]{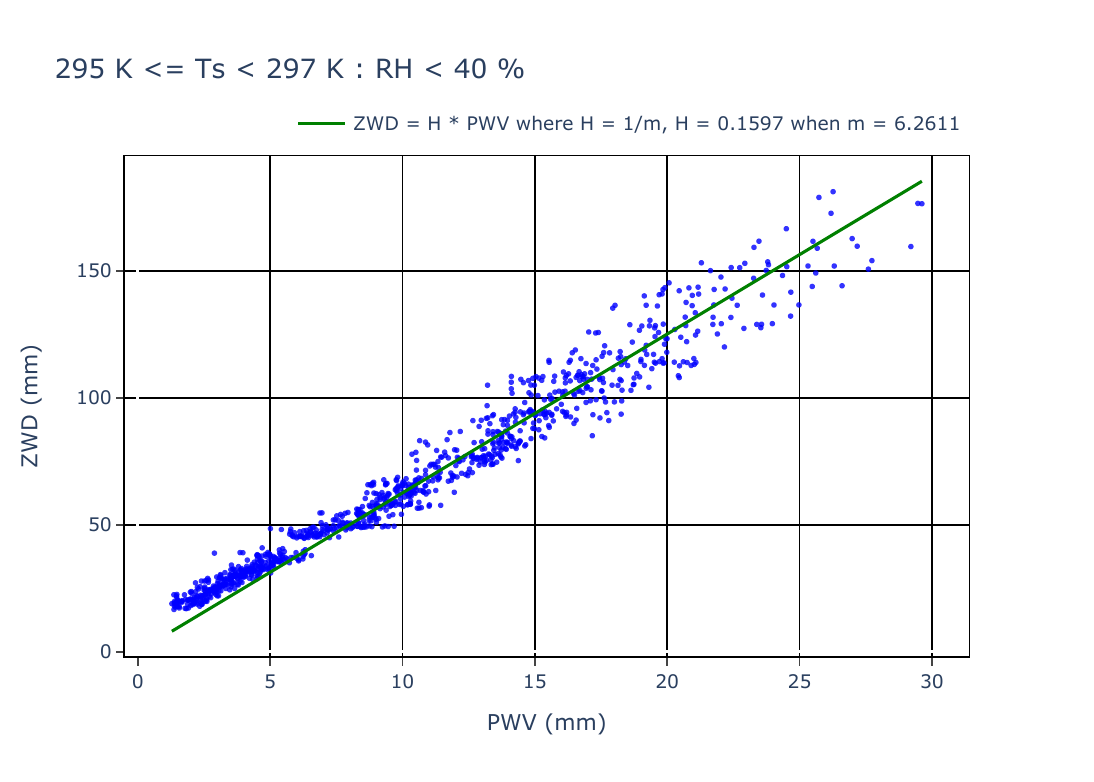}
\includegraphics[width=.32\textwidth, height =.35\textwidth]{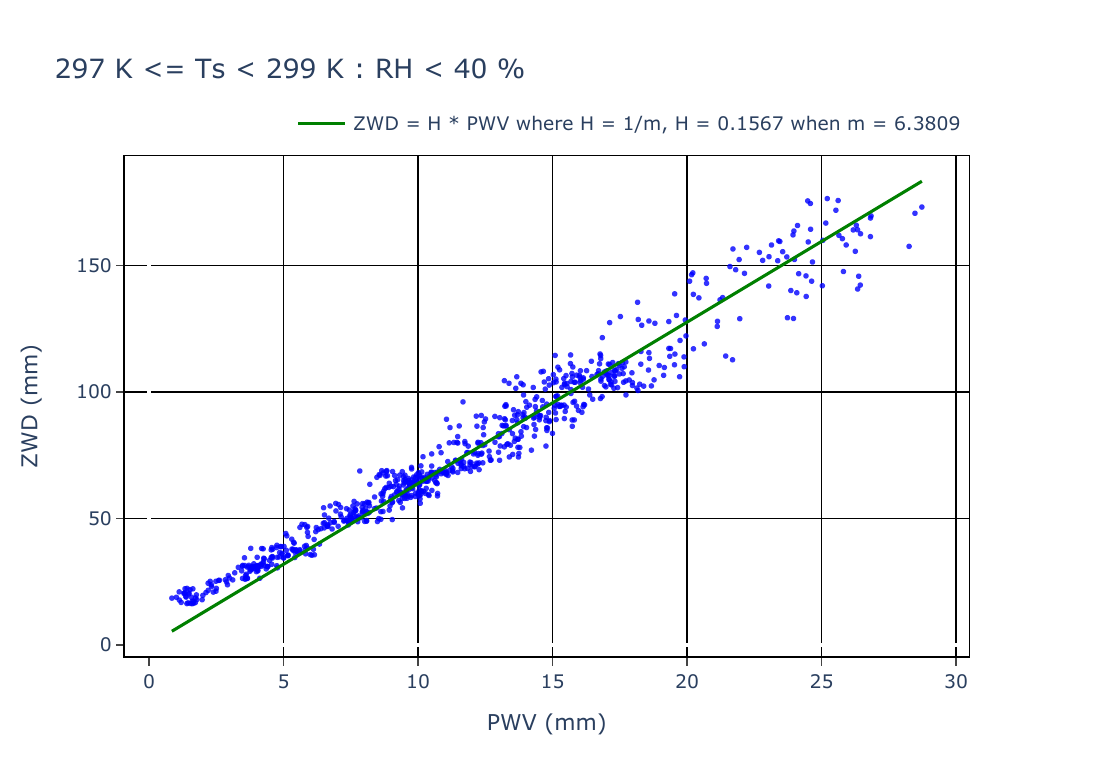}
\includegraphics[width=.32\textwidth, height =.35\textwidth]{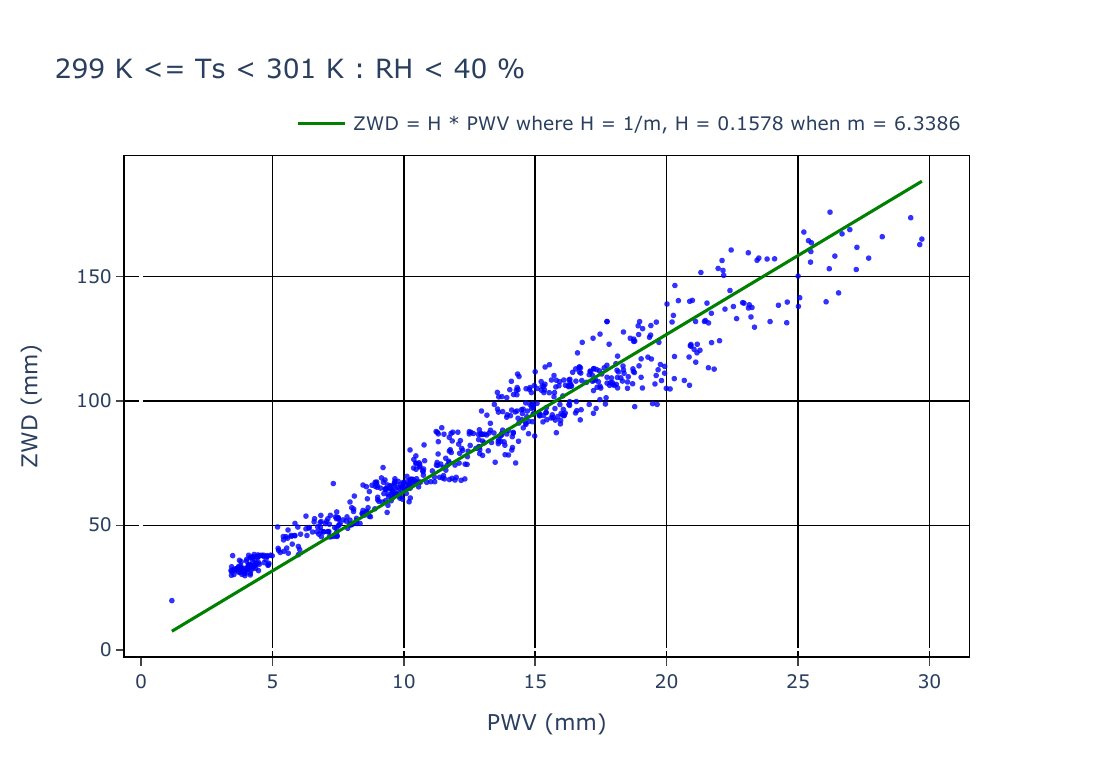}\quad

\medskip

\includegraphics[width=.32\textwidth, height =.35\textwidth]{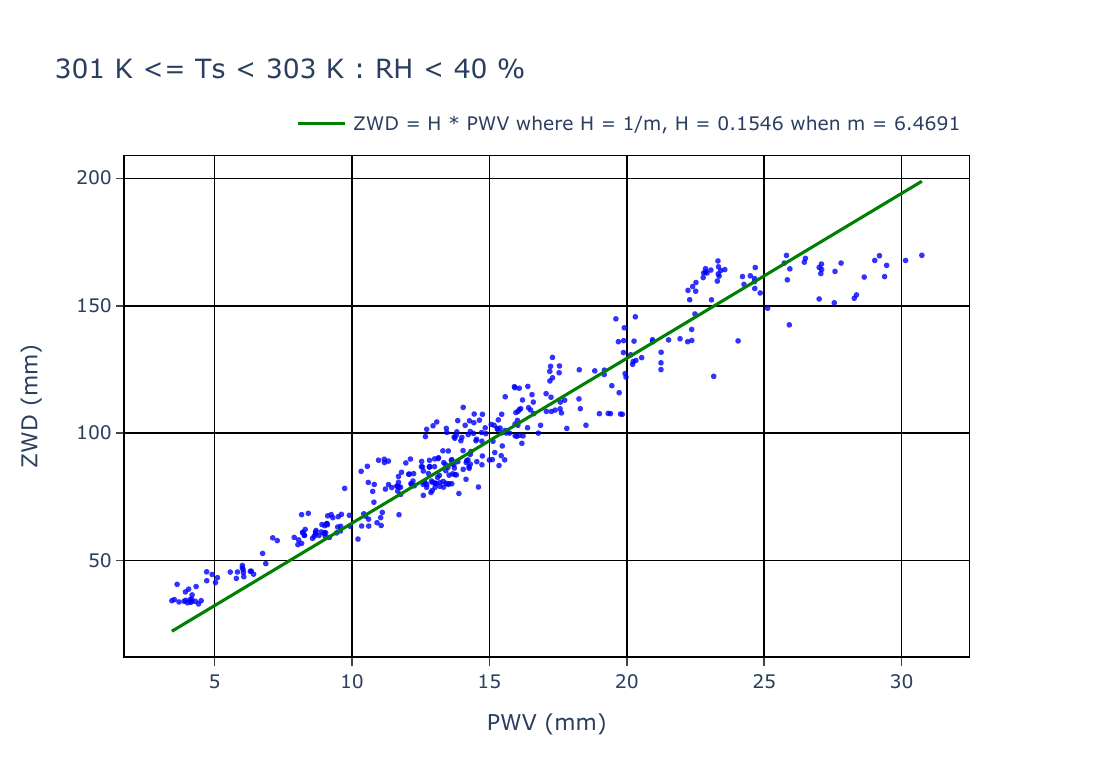}
\includegraphics[width=.32\textwidth, height =.35\textwidth]{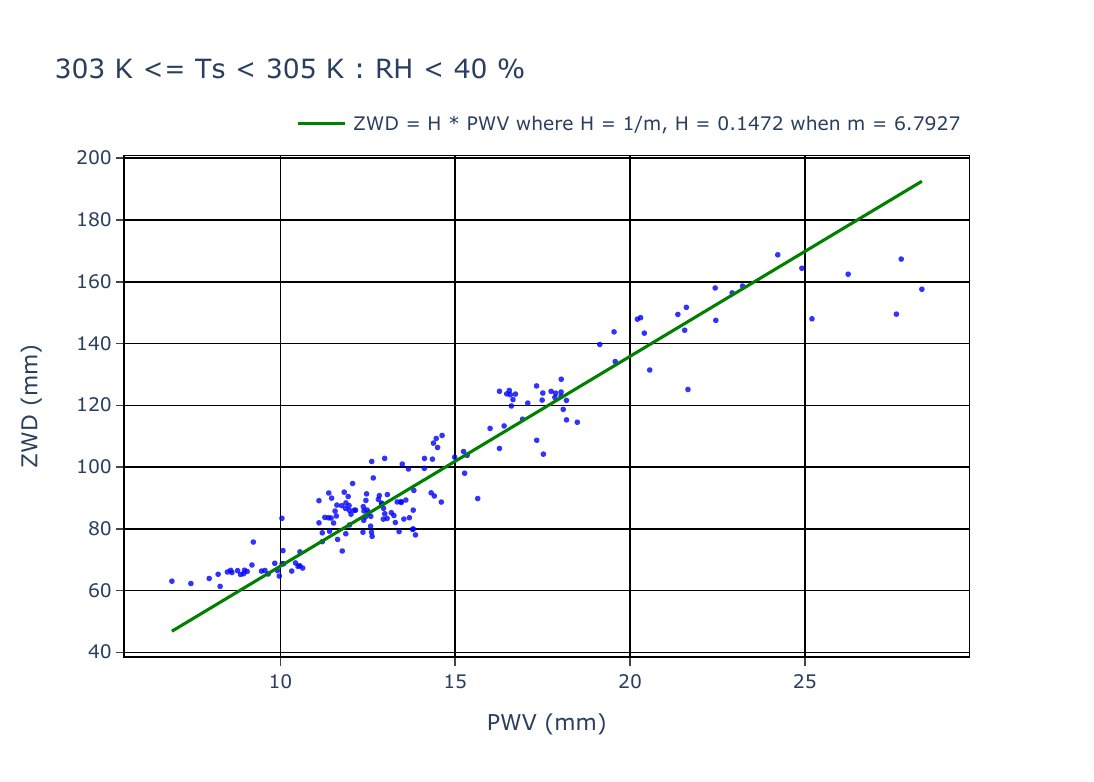}
\includegraphics[width=.32\textwidth, height =.35\textwidth]{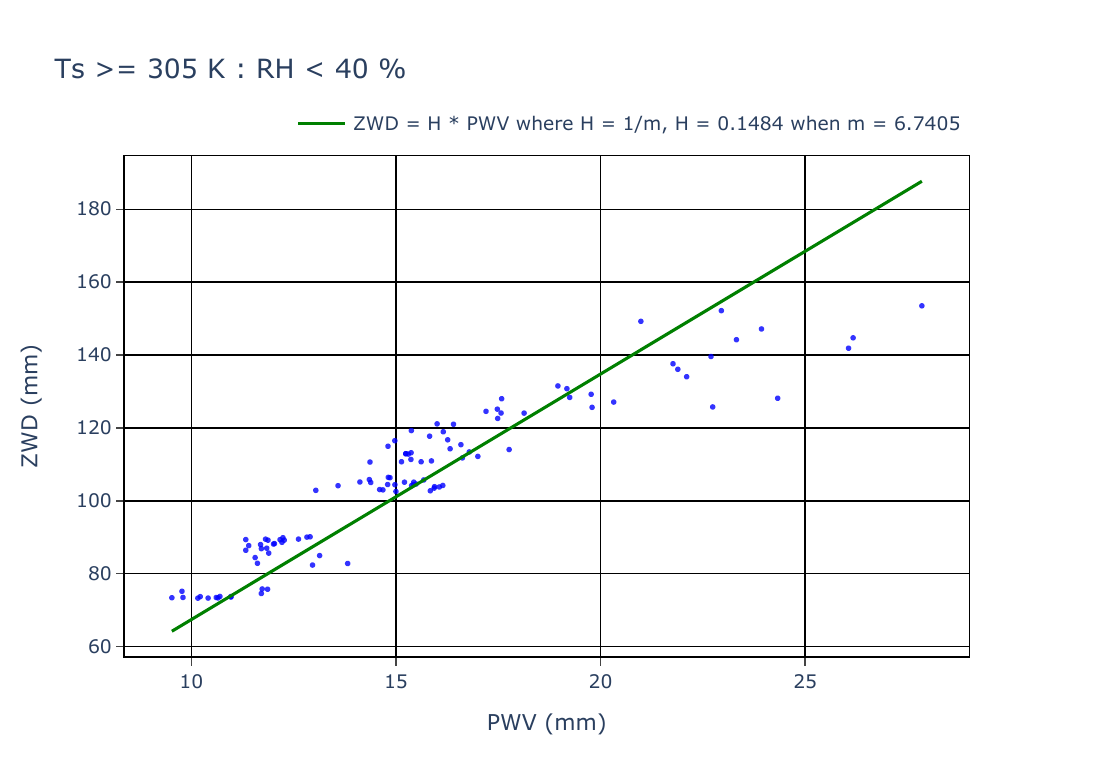}\quad

\medskip
\includegraphics[width=.32\textwidth, height =.35\textwidth]{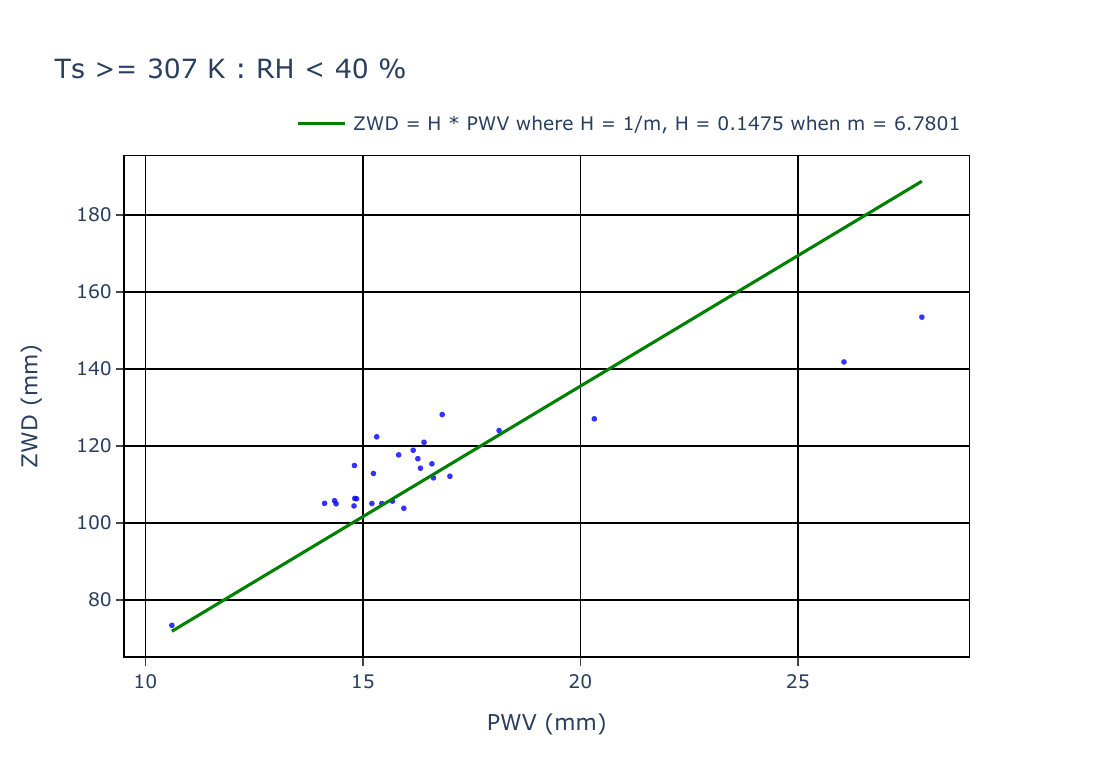}\quad

\caption{(Continued) ~ZWD against PWV at different $\mathrm{T_s}$ at the H.E.S.S. site.}
\label{fig:zwd_pwv_fit_}
\end{figure*}

\subsection{Comparison of GNSS station and 210~GHz WVR PWV data when local pressure and $\mathrm{T_s}$ are used in calculating GNSS station PWV data}
This sections shows the plot comparing the 210~GHz WVR PWV data with those of the GNSS station when insitu pressure and $\mathrm{T_s}$ are incorporated into the calculation of the PWV data of the GNSS station. The $\mathrm{T_s}$ was used to calculate the $\mathrm{T_m}$ using the model given in equation~\ref{eq:Tm_piece}. 
\begin{figure*}
    \centering
    \includegraphics[scale = 0.5]{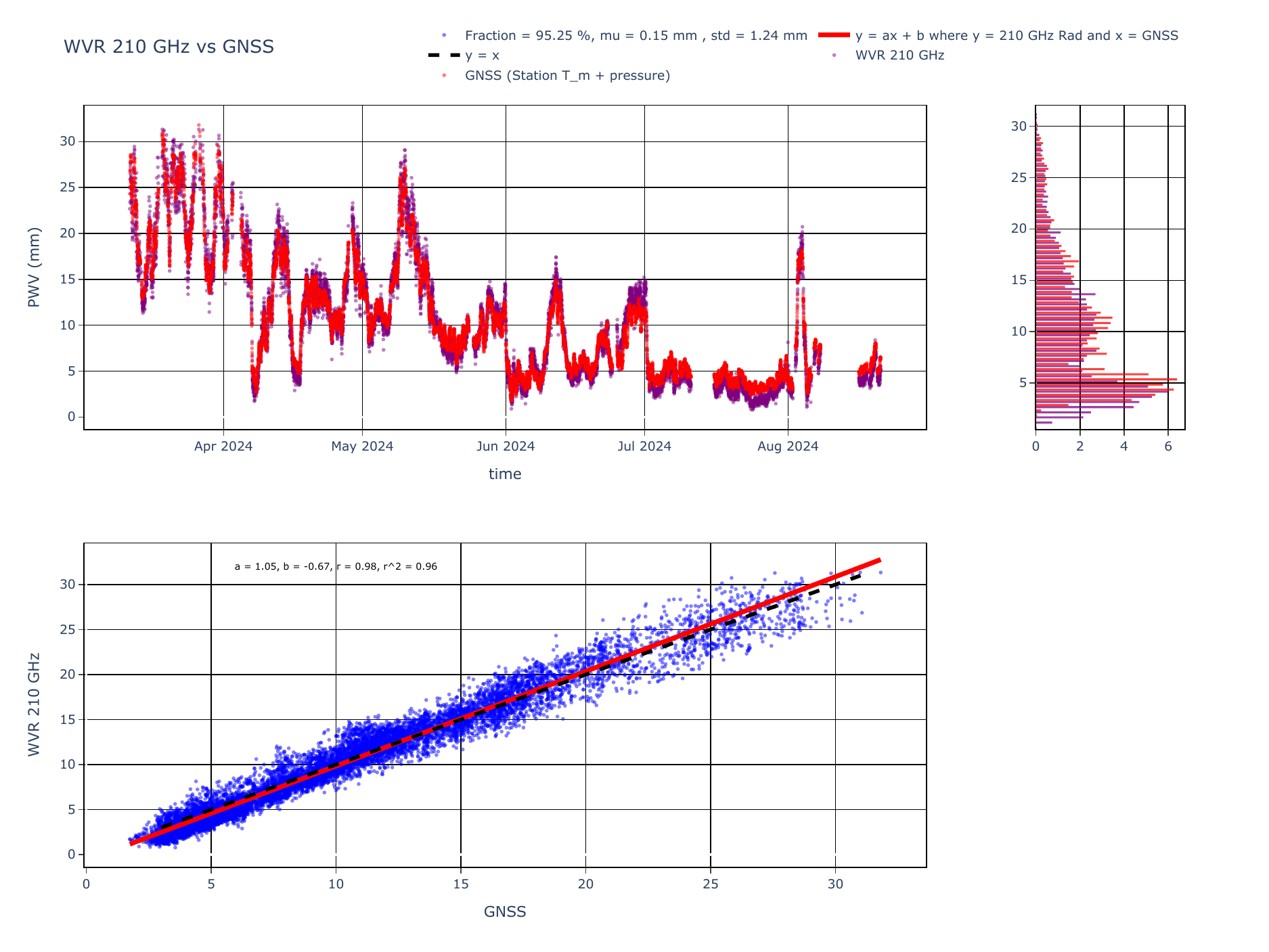}
    \caption{Time series and scatter plots of the resultant PWV from GNSS station when $\mathrm{T_m}$ model is used versus PWV from the 210~GHz WVR.}
   \label{fig:Tm_plot_results}
\end{figure*}






\bsp	
\label{lastpage}
\end{document}